\documentclass[prd,preprint,tightenlines,floatfix,showpacs,preprintnumbers,nofootinbib,eqsecnum,superscriptaddress]{revtex4}

\usepackage[dvips,final]{graphicx}
 \usepackage{amssymb}
  \usepackage{amsmath}
   \usepackage{amsfonts}
    \usepackage{epsfig}
     \usepackage{bm}

\begin{document}

\title{
Exclusive \mbox{\boldmath $\textit{\textbf{pp}} \to \textit{\textbf{nn}} \pi^{+}\pi^{+}$} reaction at LHC and RHIC}

\author{P.~Lebiedowicz}
\email{piotr.lebiedowicz@ifj.edu.pl}
\affiliation{Institute of Nuclear Physics PAN, PL-31-342 Cracow, Poland}

\author{A.~Szczurek}
\email{antoni.szczurek@ifj.edu.pl}
\affiliation{University of Rzesz\'ow, PL-35-959 Rzesz\'ow, Poland}
\affiliation{Institute of Nuclear Physics PAN, PL-31-342 Cracow, Poland}

\begin{abstract}
We evaluate differential distributions
for the four-body $p p \to n n \pi^+ \pi^+$ reaction.
The amplitude for the process is calculated in the Regge approach
including many diagrams.
We make predictions for possible future experiments at RHIC and LHC energies.
Very large cross sections are found which
is partially due to interference of a few mechanisms.
Presence of several interfering mechanisms precludes
extraction of the elastic $\pi^+ \pi^+$ scattering cross section.
Absorption effects are estimated.
Differential distributions in pseudorapidity, rapidity, invariant two-pion mass,
transverse-momentum and energy distributions of neutrons
are presented for proton-proton collisions at
$\sqrt{s}$ = 500 GeV (RHIC) and
$\sqrt{s}$ = 0.9, 2.36 and 7 TeV (LHC).
Cross sections with experimental cuts are presented.
\end{abstract}


\maketitle

\section{Introduction}

The total and elastic cross sections are basic
objects of the scattering theory. While the proton-proton,
proton-antiproton or pion-proton can be directly measured
(see e.g.\cite{PDG})
the pion-pion scattering is not directly accessible.
It was suggested recently \cite{PRS09} how to extract the total
$\pi^+ \pi^+$ cross section in the high-energy region.
Here it was suggested to use scattering of virtual $\pi^+$'s
which couple to the nucleons with well known coupling constant
and are subsequently promoted by the interaction
onto their mass shell in the final state.
The final pions are then associated with outgoing neutrons. 

Can a similar method be used
to extract the elastic $\pi^+ \pi^+$ scattering
by analysis of the $p p \to n n \pi^+ \pi^+$ reaction?
We wish to address this issue in the present paper.
\footnote{After first version of our paper
had been completed, a paper has appeared which
also discusses the possibility of extraction of elastic $\pi^{+}\pi^{+}$
cross section \cite{SRPM10}. In our analysis we take into account many 
more possible mechanisms
for the $p p \to n n \pi^+ \pi^+$ reaction.}
The energy dependence of the total and possibly elastic cross section
of pion-pion scattering would be very useful and supplementary information
for the groups which model hadron-hadron interactions in the soft sector
(see e.g. \cite{GPSS10}).

It was realized over the last decade that the measurement of
forward particles can be an interesting and useful supplement
to the central multipurpose LHC detectors (ATLAS, CMS).
The main effort concentrated on the design and construction
of forward proton detectors \cite{FP420}.
Also Zero-Degree Calorimeters (ZDC's) have been considered as
a useful supplement.
It will measure very forward neutrons and photons
in the pseudorapidity region 
$|\eta|\geq 8.5$ at the CMS \cite{ZDC_CMS} (see also \cite{Murray_Trento})
and the ATLAS ZDC's provide coverage of the region $|\eta|\geq 8.3$ \cite{ZDC_ATLAS}.
It was shown recently that the CMS (Compact Muon Spectrometer)
Collaboration ZDC's provide a unique possibility to
measure the $\pi^+ \pi^+$ total cross section \cite{PRS09}.

Even at high-energy the major part of the phase space of a few-body
reactions is populated in soft processes which cannot be calculated
within perturbative QCD. Only limited corner of the phase space,
where particles are produced at large transverse momenta, can be
addressed in the framework of pQCD.
At high energy the Regge approach is the most efficient tool
to describe total cross section, elastic scattering as well as
different 2 $\to$ 2 reactions \cite{books}.
In the present paper we shall show how to construct the amplitude
for the considered 2 $\to$ 4 process in terms of several 2 $\to$ 2
soft amplitudes known from the literature.
In the present analysis
we will also include absorption effects as was done recently
for three-body processes \cite{SS07}.

In the present paper we consider an example of an exclusive
reaction with two forward neutrons.
Given the experimental infrastructure
the $p p \to n n \pi^+ \pi^+$
is one of the reactions with four particles
in the final state which could be addressed at LHC.
\section{Amplitude of exclusive $\textit{\textbf{pp}} \to \textit{\textbf{nn}} \pi^{+}\pi^{+}$ reaction}
\subsection{Dominant diffractive amplitude}
\begin{figure}[!h]    %
a)\includegraphics[width=0.18\textwidth]{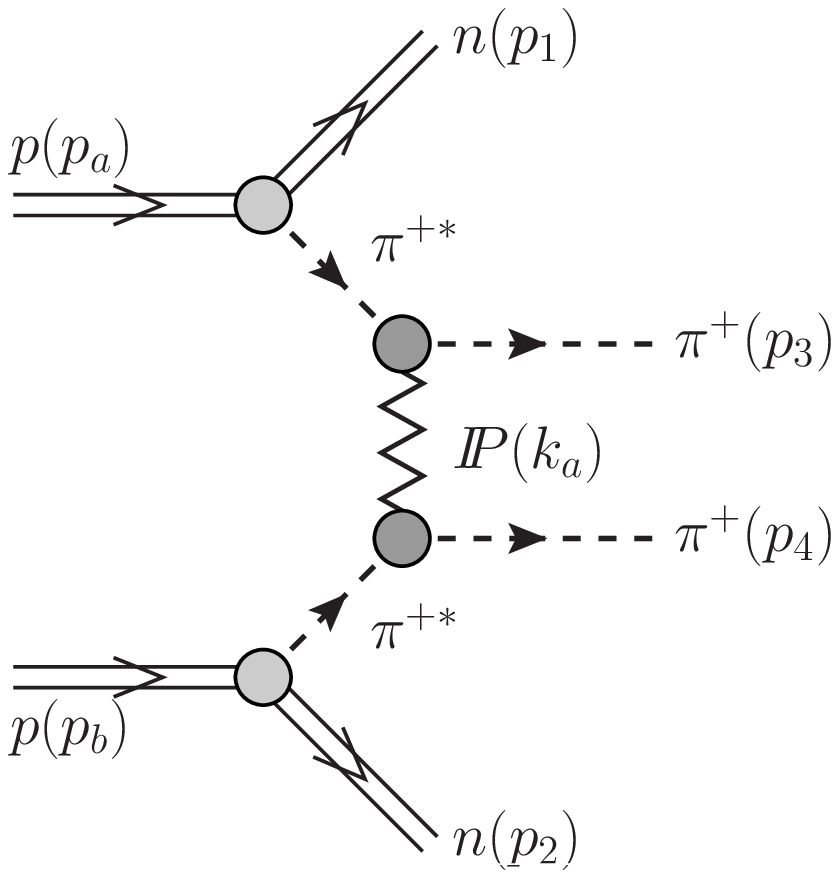}
b)\includegraphics[width=0.18\textwidth]{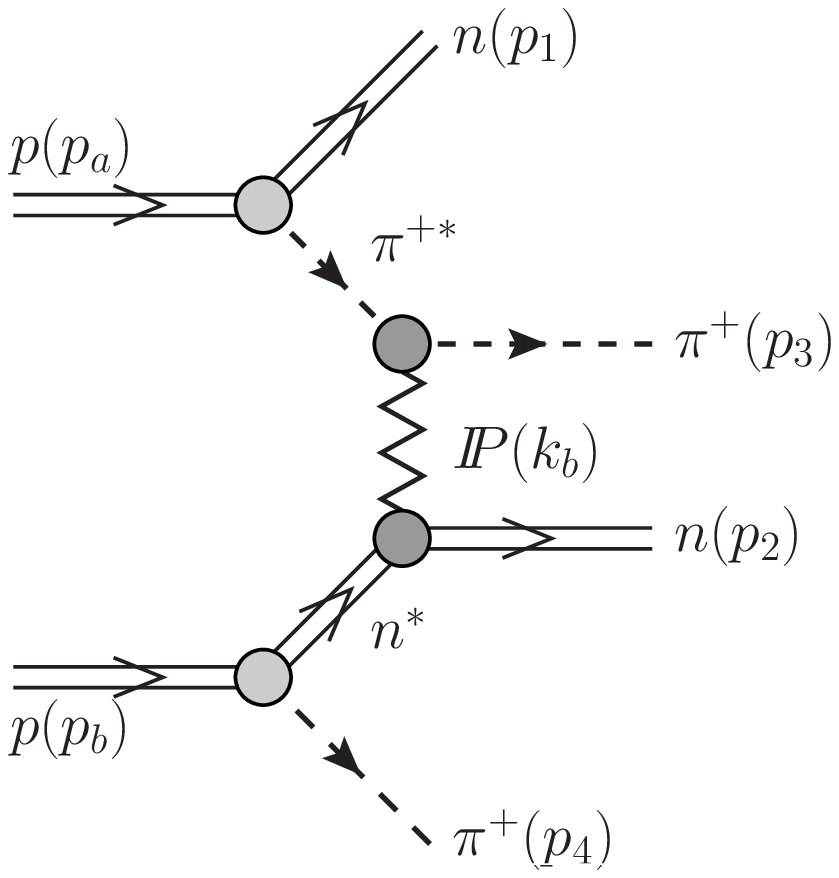}
c)\includegraphics[width=0.18\textwidth]{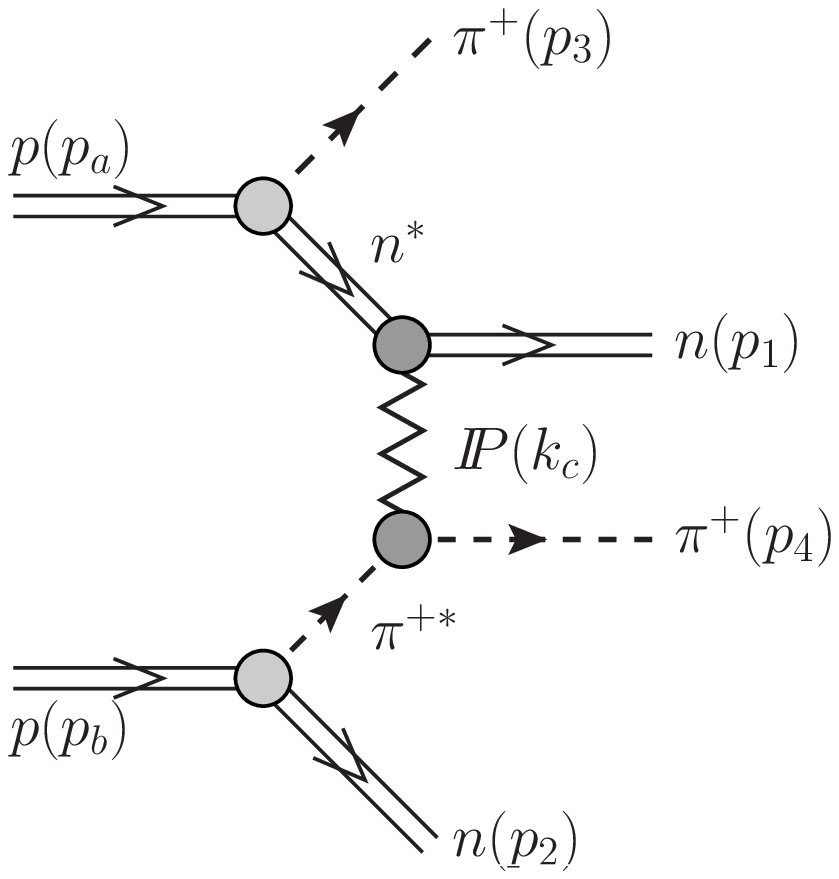}
d)\includegraphics[width=0.18\textwidth]{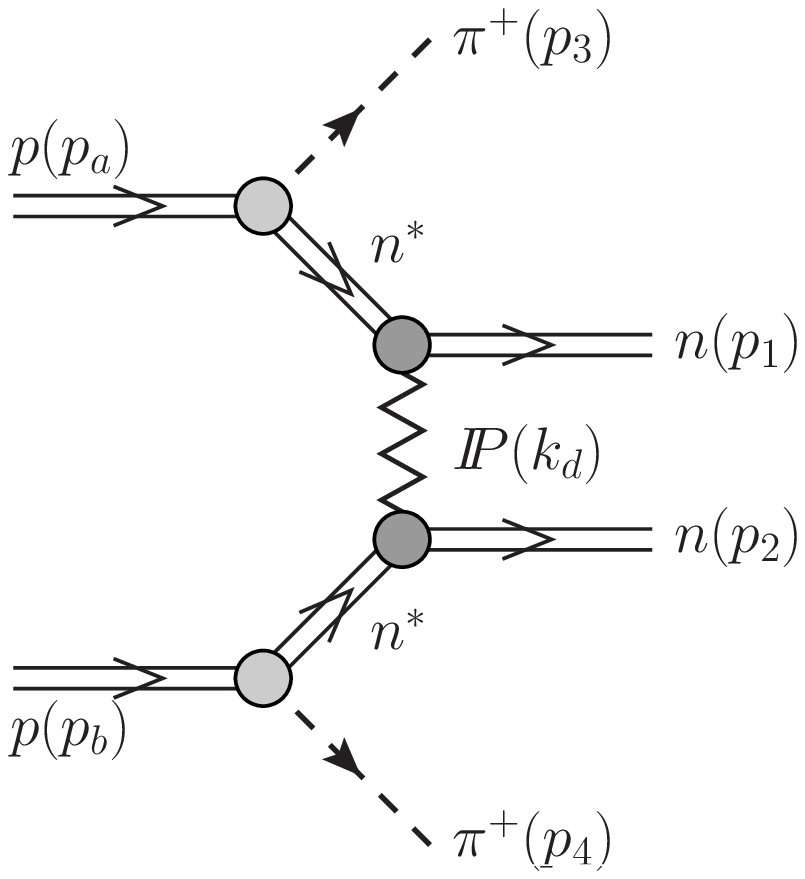}\\
e)\includegraphics[width=0.22\textwidth]{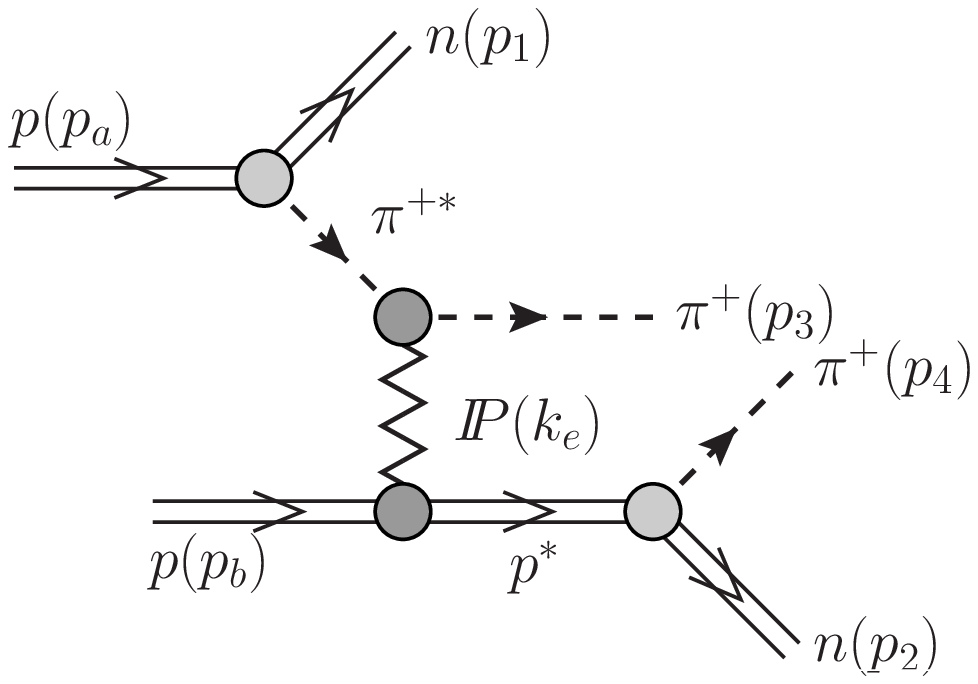} 
f)\includegraphics[width=0.22\textwidth]{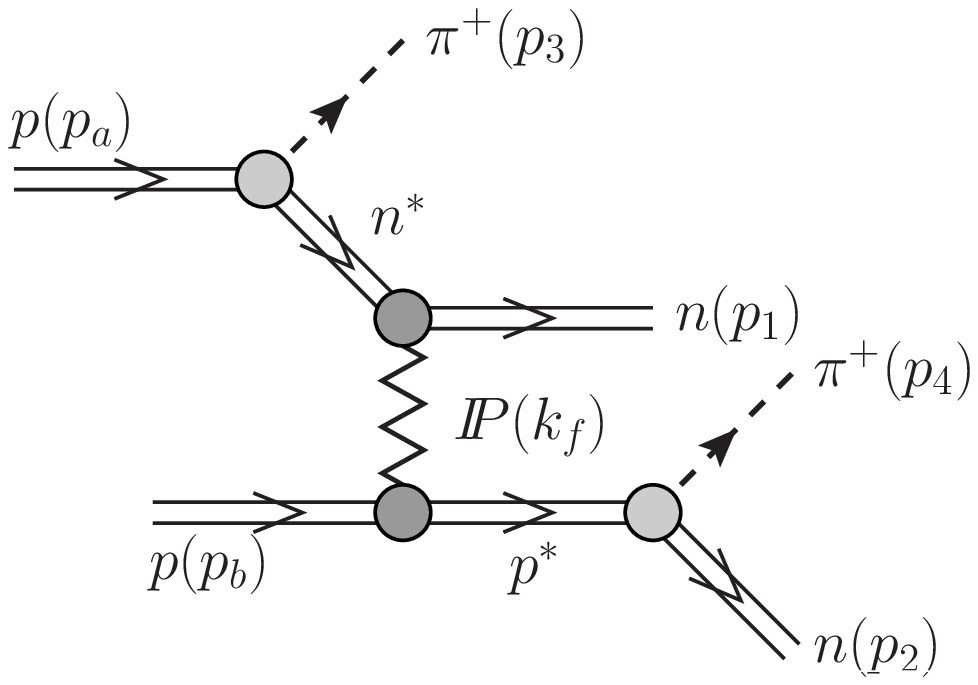}
g)\includegraphics[width=0.22\textwidth]{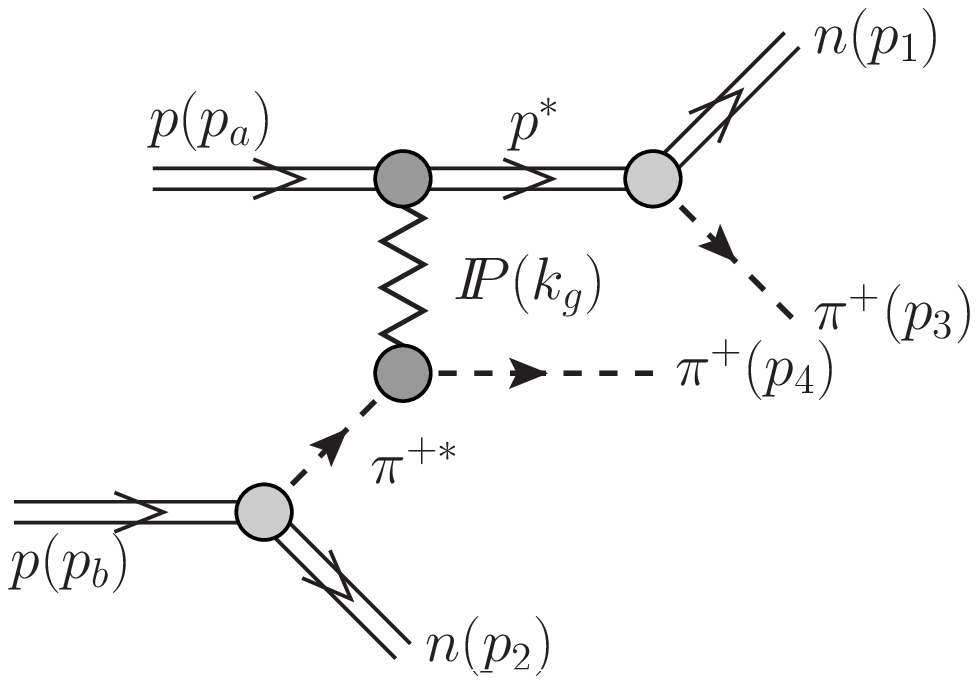} 
h)\includegraphics[width=0.22\textwidth]{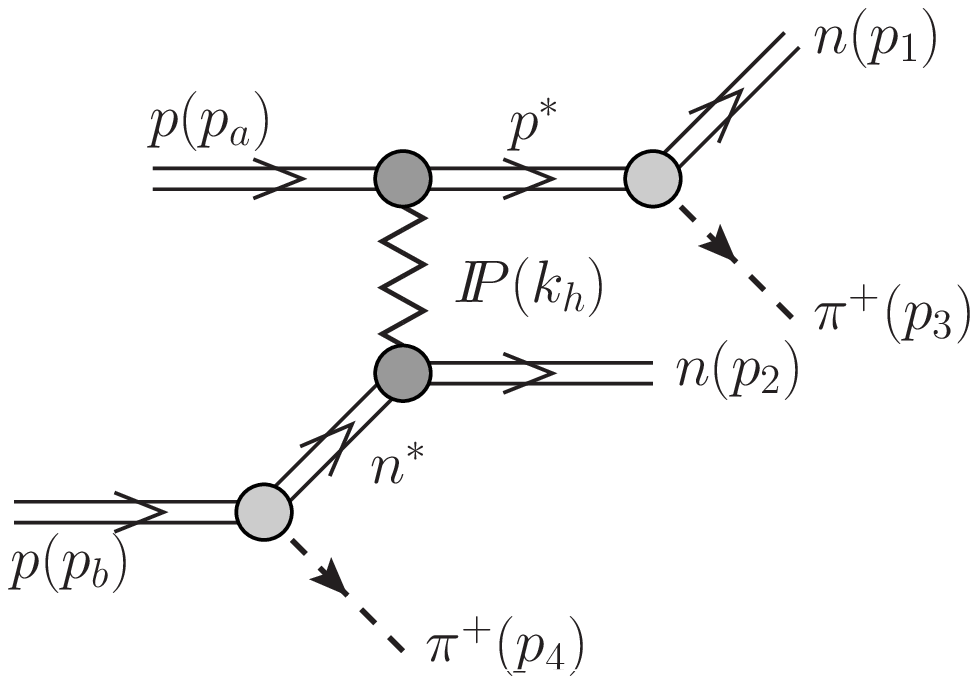}\\
i)\includegraphics[width=0.18\textwidth]{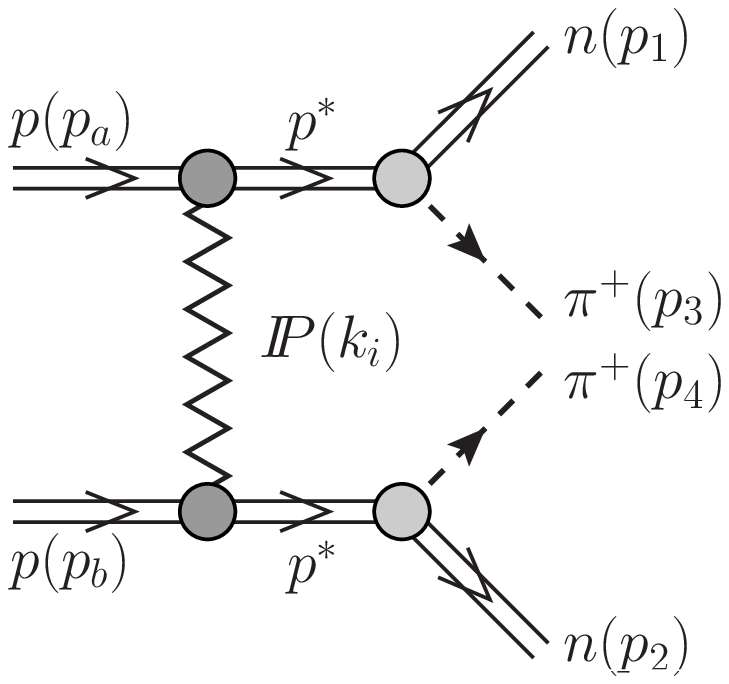} 
  \caption{\label{fig:diagrams}
  \small
Diagrams for the exclusive production of $\pi^{+}\pi^{+}$ 
in $pp$ collisions at high energies.
The stars attached to $\pi^{+}$, $n$ and $p$ denote the fact they are off-mass-shell.
$k_{a}$--$k_{i}$ are four-vectors of the exchanged pomerons.
}
\end{figure}

The diffractive mechanisms involving pomeron and reggeon exchanges
included in the present paper are shown in Fig.\ref{fig:diagrams}
(with the four-momenta $p_{a}+p_{b} \to p_{1}+p_{2}+p_{3}+p_{4}$).
In principle in all diagrams shown the intermediate nucleon
can be replaced by nucleon excited states.
It is known that diffractive excitation of nucleons to inelastic
states is rather large and constitutes about 1/3 of the elastic scattering.
This number is, however, not relevant in our case,
as it is to large extend due to the Deck type mechanism \cite{AG81}
which is included explicitly in our calculation.
The remaining excitation to discrete nucleon states
is rather small and difficult to calculate.
A microscopic calculation must unavoidably
include not only the structure of the nucleon but also
of the nucleon excited states.
The cross section for
$pp \to p + N \pi\pi$ of our interest is, however, only 
a fraction of mb \cite{HNSSZ96}.
That the contribution of excited discrete state is small
can be also seen in the following way.
First of all the diffractive transitions to discrete excited states are known to be much
weaker than the elastic one. Secondly the $g_{N N^{*} \pi}$ coupling constants
are much smaller than the $g_{N N \pi}$ coupling constant \cite{coupling_const}.
Finally the exact strength of the diffractive transitions are not known phenomenologically. 
Therefore in the following we neglect the contributions of diagrams with excited nucleon states.

Similarly as for the $p \bar{p} \to N \bar{N} f_{0}(1500)$
\cite{SL09} and $p p \to p p \pi^+ \pi^-$
($p \bar{p} \to p \bar{p} \pi^+ \pi^-$) \cite{LSK09,LS10} reactions
the amplitudes can be written in terms of pomeron (reggeon)-exchanges.
Then the amplitude squared, averaged over the initial and summed
over the final polarization states, 
for the $p p \to n n \pi^+ \pi^+$ process can be written as:
\begin{eqnarray}
\overline{|{\cal M}|^2} &=& \dfrac{1}{4} \sum_{\lambda_{a}\lambda_{b}\lambda_{1}\lambda_{2}}
|{\cal M}^{(a)}_{{\lambda_{a}\lambda_{b} \to \lambda_{1}\lambda_{2}}} + ... +
{\cal M}^{(i)}_{\lambda_{a}\lambda_{b} \to \lambda_{1}\lambda_{2}}|^2 \;.
\label{amplitude}
\end{eqnarray}
It is straightforward to evaluate
the contribution shown in
Fig.\ref{fig:diagrams}.
The diagrams will be called a) -- i) for brevity.
If we assume the $\mathrm{i} \gamma_5$ type coupling of the pion
to the nucleon then
the Born amplitudes read:
{\footnote{We show explicitly only amplitudes for pomeron exchange. 
The amplitudes for reggeon exchange can be 
obtained from those for pomeron exchange
by replacing propagators by signature factors and trajectories.}}

\begin{eqnarray}
{\cal M}^{(a)}_{\lambda_{a}\lambda_{b} \to \lambda_{1}\lambda_{2}} &=&
\bar{u}(p_{1},\lambda_{1}) \mathrm{i} \gamma_{5}  S_{\pi}(t_{1}) u(p_{a},\lambda_{a})
\sqrt{2} g_{\pi NN} F_{\pi NN}(t_{1}) \nonumber \\
&\times &
F_{\pi}^{off}(t_{1}) \;
\mathrm{i} s_{34} C_{I\!\!P}^{\pi\pi} \left( \frac{s_{34}}{s_{0}}\right)^{\alpha_{I\!\!P}(k_{a}^{2})-1}
\exp \left( {\frac{B_{\pi \pi}}{2} k_{a}^{2} }\right) \;
F_{\pi}^{off}(t_{2}) \nonumber \\
&\times &
\bar{u}(p_{2},\lambda_{2}) \mathrm{i} \gamma_{5} S_{\pi}(t_{2}) u(p_{b},\lambda_{b})
\sqrt{2} g_{\pi NN} F_{\pi NN}(t_{2}) \;,
\label{amp_a}
\end{eqnarray}
\begin{eqnarray}
{\cal M}^{(b)}_{\lambda_{a}\lambda_{b} \to \lambda_{1}\lambda_{2}} &=&
\bar{u}(p_{1},\lambda_{1}) \mathrm{i} \gamma_{5} S_{\pi}(t_{1}) u(p_{a},\lambda_{a})
\sqrt{2} g_{\pi NN} F_{\pi NN}(t_{1}) \nonumber \\
&\times &
F_{\pi}^{off}(t_{1}) \;
\mathrm{i} s_{23} C_{I\!\!P}^{\pi N} \left( \frac{s_{23}}{s_{0}}\right)^{\alpha_{I\!\!P}(k_{b}^{2})-1}
\left( \frac{s_{24}}{s_{th}}\right)^{\alpha_{N}(u_{2})-\frac{1}{2}}
\exp \left( {\frac{B_{\pi N}}{2} k_{b}^{2} }\right) \;
F_{n}^{off}(u_{2})\nonumber \\
&\times &
\bar{u}(p_{2},\lambda_{2}) \mathrm{i} \gamma_{5} S_{n}(u_{2}) u(p_{b},\lambda_{b})
\sqrt{2} g_{\pi NN} F_{\pi NN}(u_{2}) \;,
\label{amp_b}
\end{eqnarray}
\begin{eqnarray}
{\cal M}^{(c)}_{\lambda_{a}\lambda_{b} \to \lambda_{1}\lambda_{2}} &=&
\bar{u}(p_{1},\lambda_{1}) \mathrm{i} \gamma_{5} S_{n}(u_{1}) u(p_{a},\lambda_{a})
\sqrt{2} g_{\pi NN} F_{\pi NN}(u_{1}) \nonumber \\
&\times &
F_{n}^{off}(u_{1}) \;
\mathrm{i} s_{14} C_{I\!\!P}^{\pi N} \left( \frac{s_{14}}{s_{0}}\right)^{\alpha_{I\!\!P}(k_{c}^{2})-1}
\left( \frac{s_{13}}{s_{th}}\right)^{\alpha_{N}(u_{1})-\frac{1}{2}}
\exp \left( {\frac{B_{\pi N}}{2} k_{c}^{2}}\right) \;
F_{\pi}^{off}(t_{2}) \nonumber \\
&\times &
\bar{u}(p_{2},\lambda_{2}) \mathrm{i} \gamma_{5} S_{\pi}(t_{2}) u(p_{b},\lambda_{b})
\sqrt{2} g_{\pi NN} F_{\pi NN}(t_{2}) \;,
\label{amp_c}
\end{eqnarray}
\begin{eqnarray}
{\cal M}^{(d)}_{\lambda_{a}\lambda_{b} \to \lambda_{1}\lambda_{2}} &=&
\bar{u}(p_{1},\lambda_{1}) \mathrm{i} \gamma_{5} S_{n}(u_{1}) u(p_{a},\lambda_{a})
\sqrt{2} g_{\pi NN} F_{\pi NN}(u_{1}) \nonumber \\
&\times &
F_{n}^{off}(u_{1}) \;
\mathrm{i} s_{12} C_{I\!\!P}^{NN} \left( \frac{s_{12}}{s_{0}}\right)^{\alpha_{I\!\!P}(k_{d}^{2})-1}
\left( \frac{s_{13}}{s_{th}}\right)^{\alpha_{N}(u_{1})-\frac{1}{2}}
\left( \frac{s_{24}}{s_{th}}\right)^{\alpha_{N}(u_{2})-\frac{1}{2}}\nonumber \\
&\times &
\exp \left( {\frac{B_{NN}}{2} k_{d}^{2}}\right) \;
F_{n}^{off}(u_{2}) \nonumber \\
&\times &
\bar{u}(p_{2},\lambda_{2}) \mathrm{i} \gamma_{5} S_{n}(u_{2}) u(p_{b},\lambda_{b})
\sqrt{2} g_{\pi NN} F_{\pi NN}(u_{2}) \;,
\label{amp_d}
\end{eqnarray}
\begin{eqnarray}
{\cal M}^{(e)}_{\lambda_{a}\lambda_{b} \to \lambda_{1}\lambda_{2}} &=&
\bar{u}(p_{1},\lambda_{1}) \mathrm{i} \gamma_{5} S_{\pi}(t_{1}) u(p_{a},\lambda_{a})
\sqrt{2} g_{\pi NN} F_{\pi NN}(t_{1}) \nonumber \\
&\times &
F_{\pi}^{off}(t_{1}) \;
\mathrm{i} s_{234} C_{I\!\!P}^{\pi N} \left( \frac{s_{234}}{s_{0}}\right)^{\alpha_{I\!\!P}(k_{e}^{2})-1}
\exp \left( {\frac{B_{\pi N}}{2} k_{e}^{2}}\right) \;
F_{p}^{off}(s_{24}) \nonumber \\
&\times &
\bar{u}(p_{2},\lambda_{2}) \mathrm{i} \gamma_{5} S_{p}(s_{24}) u(p_{b},\lambda_{b})
\sqrt{2} g_{\pi NN} F_{\pi NN}(s_{24}) \;,
\label{amp_e}
\end{eqnarray}
\begin{eqnarray}
{\cal M}^{(f)}_{\lambda_{a}\lambda_{b} \to \lambda_{1}\lambda_{2}} &=&
\bar{u}(p_{1},\lambda_{1}) \mathrm{i} \gamma_{5} S_{n}(u_{1}) u(p_{a},\lambda_{a})
\sqrt{2} g_{\pi NN} F_{\pi NN}(u_{1}) \nonumber \\
&\times &
F_{n}^{off}(u_{1}) \;
\mathrm{i} s_{124} C_{I\!\!P}^{NN} \left( \frac{s_{124}}{s_{0}}\right)^{\alpha_{I\!\!P}(k_{f}^{2})-1}
\left( \frac{s_{13}}{s_{th}}\right)^{\alpha_{N}(u_{1})-\frac{1}{2}}
\exp \left( {\frac{B_{NN}}{2} k_{f}^{2}}\right) \;
F_{p}^{off}(s_{24}) \nonumber \\
&\times &
\bar{u}(p_{2},\lambda_{2}) \mathrm{i} \gamma_{5} S_{p}(s_{24}) u(p_{b},\lambda_{b})
\sqrt{2} g_{\pi NN} F_{\pi NN}(s_{24}) \;,
\label{amp_f}
\end{eqnarray}
\begin{eqnarray}
{\cal M}^{(g)}_{\lambda_{a}\lambda_{b} \to \lambda_{1}\lambda_{2}} &=&
\bar{u}(p_{1},\lambda_{1}) \mathrm{i} \gamma_{5} S_{p}(s_{13}) u(p_{a},\lambda_{a})
\sqrt{2} g_{\pi NN} F_{\pi NN}(s_{13}) \nonumber \\
&\times &
F_{p}^{off}(s_{13}) \;
\mathrm{i} s_{134} C_{I\!\!P}^{\pi N} \left( \frac{s_{134}}{s_{0}}\right)^{\alpha_{I\!\!P}(k_{g}^{2})-1}
\exp \left( {\frac{B_{\pi N}}{2} k_{g}^{2}}\right) \;
F_{\pi}^{off}(t_{2}) \nonumber \\
&\times &
\bar{u}(p_{2},\lambda_{2}) \mathrm{i} \gamma_{5} S_{\pi}(t_{2}) u(p_{b},\lambda_{b})
\sqrt{2} g_{\pi NN} F_{\pi NN}(t_{2}) \;,
\label{amp_g}
\end{eqnarray}
\begin{eqnarray}
{\cal M}^{(h)}_{\lambda_{a}\lambda_{b} \to \lambda_{1}\lambda_{2}} &=&
\bar{u}(p_{1},\lambda_{1}) \mathrm{i} \gamma_{5} S_{p}(s_{13}) u(p_{a},\lambda_{a})
\sqrt{2} g_{\pi NN} F_{\pi NN}(s_{13}) \nonumber \\
&\times &
F_{p}^{off}(s_{13}) \;
\mathrm{i} s_{123} C_{I\!\!P}^{NN} \left( \frac{s_{123}}{s_{0}}\right)^{\alpha_{I\!\!P}(k_{h}^{2})-1}
\left( \frac{s_{24}}{s_{th}}\right)^{\alpha_{N}(u_{2})-\frac{1}{2}}
\exp \left( {\frac{B_{NN}}{2} k_{h}^{2}}\right) \;
F_{n}^{off}(u_{2}) \nonumber \\
&\times &
\bar{u}(p_{2},\lambda_{2}) \mathrm{i} \gamma_{5} S_{n}(u_{2}) u(p_{b},\lambda_{b})
\sqrt{2} g_{\pi NN} F_{\pi NN}(u_{2}) \;,
\label{amp_h}
\end{eqnarray}
\begin{eqnarray}
{\cal M}^{(i)}_{\lambda_{a}\lambda_{b} \to \lambda_{1}\lambda_{2}} &=&
\bar{u}(p_{1},\lambda_{1}) \mathrm{i} \gamma_{5} S_{p}(s_{13}) u(p_{a},\lambda_{a})
\sqrt{2} g_{\pi NN} F_{\pi NN}(s_{13}) \nonumber \\
&\times &
F_{p}^{off}(s_{13}) \;
\mathrm{i} s_{ab} C_{I\!\!P}^{NN} \left( \frac{s_{ab}}{s_{0}}\right)^{\alpha_{I\!\!P}(k_{i}^{2})-1}
\exp \left( {\frac{B_{NN}}{2} k_{i}^{2}}\right) \;
F_{p}^{off}(s_{24}) \nonumber \\
&\times &
\bar{u}(p_{2},\lambda_{2}) \mathrm{i} \gamma_{5} S_{p}(s_{24}) u(p_{b},\lambda_{b})
\sqrt{2} g_{\pi NN} F_{\pi NN}(s_{24}) \;,
\label{amp_i}
\end{eqnarray} 
where the energy scale $s_{0}$ is fixed at $s_{0} = 1$ GeV$^2$
and $s_{th} = (m_{N}+m_{\pi})^2$.

In the above equations
$u(p_{i},\lambda_{i})$, $\bar{u}(p_{f},\lambda_{f})=u^{\dagger}(p_{f},\lambda_{f})\gamma^{0}$
are the Dirac spinors (normalized as $\bar{u}(p) u(p) = 2 m_{N}$) 
of the initial protons and outgoing neutrons
with the four-momentum $p$ and the helicities of the nucleons $\lambda$.
The propagators of virtual particles can be written as
\begin{eqnarray}
S_{\pi}(t_{1,2})&=& {\frac{\mathrm{i}}{t_{1,2} - m_{\pi}^{2}}}\;,\\
S_{n}(u_{1,2}) &=& {\frac{\mathrm{i}(\tilde{u}_{1,2_{\nu}} \gamma^{\nu} + m_{n})}{u_{1,2} - m_{n}^{2}}}\;,\\
S_{p}(s_{ij}) &=& {\frac{\mathrm{i}(\tilde{s}_{ij_{\nu}} \gamma^{\nu} + m_{p})}{s_{ij} - m_{p}^{2}}}\;,
\label{propagators}
\end{eqnarray}
where $t_{1,2}=(p_{a,b}-p_{1,2})^{2}$ and
$u_{1,2}=(p_{a,b}-p_{3,4})^{2}=\tilde{u}^{2}_{1,2}$
are the four-momenta squared of transferred pions and neutrons, respectively
\footnote{In the following for brevity we shall
use notation $t_{1,2}$ which means $t_{1}$ or $t_{2}$.}.
$s_{ij}=(p_{i}+p_{j})^{2}=\tilde{s}^{2}_{ij}$
are the squared invariant masses of the $(i,j)$ system,
$m_{\pi}$ and $m_{n}$, $m_{p}$ are the pion and nucleons masses, respectively.
The factor $g_{\pi NN}$ is the pion nucleon coupling constant
which is relatively well known \cite{ELT02} ($g^2_{\pi N N}/4\pi$ = 13.5 -- 14.6).
In our calculations the coupling constant is taken as
$g^2_{\pi N N}/4\pi$ = 13.5.

Using the known strength parameters for the $NN$ and $\pi N$ scattering
fitted to the corresponding total cross sections
(the Donnachie-Landshoff model \cite{DL92}) we obtain
$C_{I\!\!P}^{NN}$,
$C_{I\!\!P}^{\pi N}$  
and assuming Regge factorization \cite{SNS02}
$C_{I\!\!P}^{\pi \pi}$.
The pomeron/reggeon trajectories
determined from elastic and total cross sections
are given in the linear approximation
\footnote{For simplicity we use the linear pomeron/reggeons trajectories, but further
improvements are possible.}
($\alpha_{i}(t)=\alpha_{i}(0)+\alpha_{i}^{'}t$)
%
%
where the values of relevant parameters
(the intercept $\alpha_{i}(0)$ and
the slope of trajectory $\alpha_{i}^{'}$ in GeV$^{-2}$) are also
taken from the Donnachie-Landshoff model \cite{DL92} for consistency.
Parameters of reggeon exchanges used in the present calculations
are listed in Table \ref{tab:parameters}.

The slope parameter can be writen as
\begin{eqnarray}
B(s) = B_{0} + 2 \alpha^{'}_{I\!\!P} \ln \left( \frac{s}{s_{0}}\right)\;,
\label{slope}
\end{eqnarray}
where $B_{0}$ is the $t$-slope of the elastic differential cross section.
In our calculation we use $B_{0}$:
$B_{\pi N}$ = 6.5 GeV$^{-2}$, 
$B_{NN}$ = 9 GeV$^{-2}$ and
$B_{\pi\pi}$ = 4 GeV$^{-2}$.
The value of $B_{\pi\pi}$ is not well known, however
the Regge factorization entails
$B_{\pi\pi} \approx 2 B_{\pi N} - B_{NN}$ \cite{SNS02}.
We have parametrized the
$k_{a}^{2},..., k_{i}^{2}$ dependences in the exponential form
(see formulas (\ref{amp_a}) -- (\ref{amp_i})).

We improve the parametrization of the amplitudes
for neutron exchange
(\ref{amp_b}, \ref{amp_c}, \ref{amp_d}, \ref{amp_f}, \ref{amp_h})
by the factors
$\left( \frac{s_{ij}}{s_{th}} \right)^{\alpha_{N}(u_{1,2})-\frac{1}{2}}$
to reproduce the high-energy Regge dependence.
The degenerate nucleon trajectory is
$\alpha_{N}(u_{1,2})=-0.3 + \alpha'_{N} \, u_{1,2}$, with $\alpha'_{N}=0.9$ GeV$^{-2}$.

The extra correction factors $F_{\pi, N}^{off}(k^{2})$
(where $k^2 = t_{1,2}, u_{1,2}, s_{ij}$)
are due to off-shellness of particles.
In the case of our 4-body reaction
rather large transferred four-momenta squared $k^{2}$ are involved
and one has to include non-point-like and
off-shellness nature of the particles
involved in corresponding vertices.
This is incorporated via $F_{\pi NN}(k^{2})$ vertex form factors.
We parametrize these form factors in the following exponential form:
%
\begin{eqnarray}
F(t_{1,2})&=&\exp\left(\frac{t_{1,2}-m_{\pi}^{2}}{\Lambda^{2}}\right)\;,\\
F(u_{1,2})&=&\exp\left(\frac{u_{1,2}-m_{n}^{2}}{\Lambda^{2}}\right)\;,\\
F(s_{ij})&=&\exp\left(\frac{-(s_{ij}-m_{p}^{2})}{\Lambda^{2}}\right)\;.
\label{off-shell_form_factors}
\end{eqnarray}
%
While four-momenta squared of transferred pions $t_{1,2} <0$, it is not the case for
transferred neutrons where $u_{1,2} < m_n^2$.
In general, the cut-off parameter $\Lambda_{off}$ is not known
but in principle could be fitted to the
(normalized) experimental data.
From our general experience in hadronic physics
we expect $\Lambda_{off}\sim$ 1 GeV.
Typical values of the $\pi NN$ form factor parameters
used in the meson exchange models
are $\Lambda$ = 1.2--1.4 GeV \cite{MHE87},
however the Gottfried Sum Rule violation prefers smaller
$\Lambda \approx$ 0.8 GeV \cite{GSR}.
In our calculation, if not otherwise mentioned,
we use $\Lambda = \Lambda_{off}$ = 1 GeV.
We shall discuss how uncertainties
of the form factors influence our final results.  

\begin{table}
\caption{Parameters of reggeon exchanges used in the present calculations.
}
\label{tab:parameters}
\begin{center}
\begin{tabular}{|c||c|c|c|c|c|c|}
\hline
       i & $\eta_{i}$               & $\alpha_{i}(t)$   & $C_{i}^{N N}$ (mb) & $C_{i}^{\pi N}$ (mb) & $C_{i}^{\pi \pi}$ (mb) & $r_{T}^{i}$\\
\hline 
$I\!\!P$ & $\mathrm{i}$             & 1.0808 + (0.25 GeV$^{-2}$) $t$ & 21.7    & 13.63              & 8.56                   & $-$\\ 
$f_{2}$ & $(-0.860895+\mathrm{i})$  & 0.5475 + (0.93 GeV$^{-2}$)  $t$ & 75.4875 & 31.79             & $\simeq$ 13.39         & $-$\\ 
$\rho$   & $(-1.16158 - \mathrm{i})$& 0.5475 + (0.93 GeV$^{-2}$) $t$ &  1.0925 &  4.23              & $\simeq$ 16.38         & 7.5\\ 
$a_{2}$  & $(-1 + \mathrm{i})$      & 0.5    + (0.9 GeV$^{-2}$)  $t$ &  1.7475 & $-$                & $-$                    & 6\\
$\omega$ & $(-1 - \mathrm{i})$      & 0.5    + (0.9 GeV$^{-2}$)  $t$ & 20.0625 & $-$                & $-$                    & 0\\
\hline
\end{tabular}
\end{center}
\end{table}
\subsection{Single and double charge exchanges with subleading reggeons $\rho^{+}$, $a_{2}^{+}$}
\label{subsection:charge_exchange}

\begin{figure}[!h]    %
a) \includegraphics[width=0.18\textwidth]{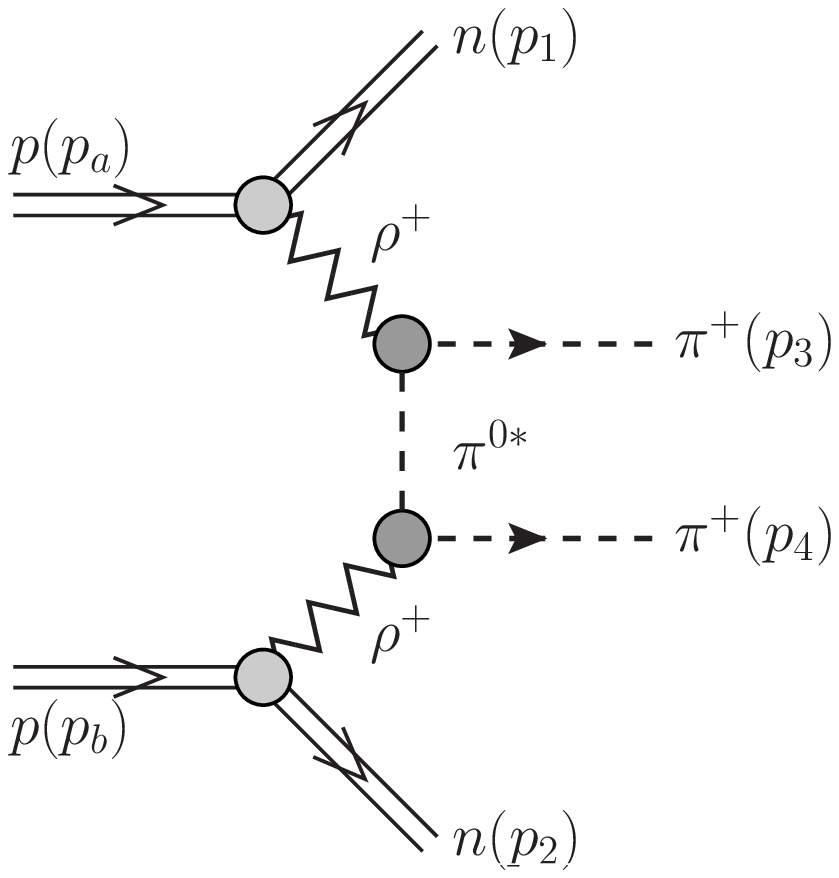}
b) \includegraphics[width=0.18\textwidth]{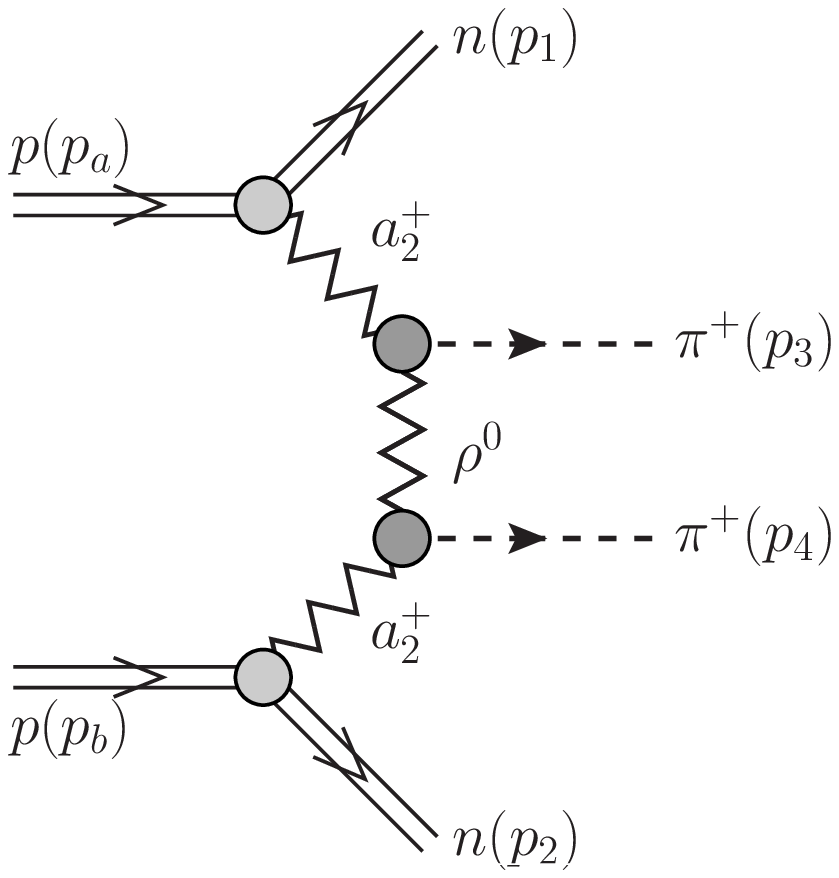}
c) \includegraphics[width=0.18\textwidth]{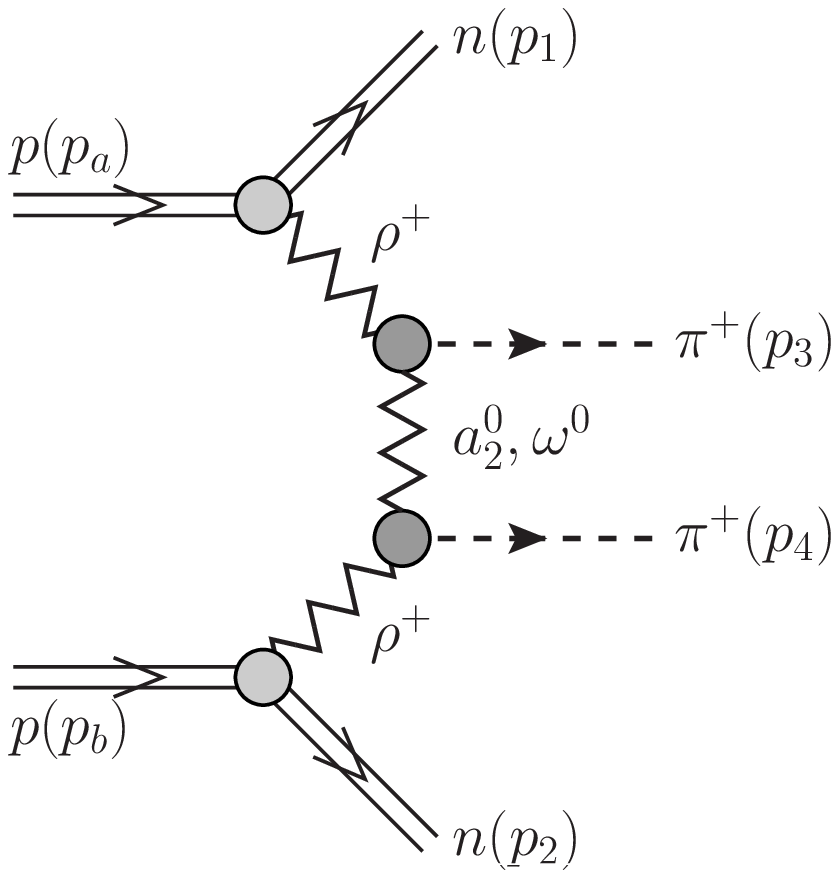}
  \caption{\label{fig:diagrams2}
  \small
Diagrams with subleading charged reggeon exchanges 
in $pp$ collisions at high energies.
}
\end{figure}

We wish to include also specific processes with 
isovector reggeon exchanges.
We include processes shown in Fig.\ref{fig:diagrams2}.
These processes involve
$\rho^+ \rho^+ \to \pi^+ \pi^+$ and $a_2^+ a_2^+ \to \pi^+ \pi^+$ subprocesses.
Unfortunately these subprocesses (or the reverse ones)
could not be studied experimentally.

The relevant coupling constants in diagrams b) and c) 
are not known and cannot be obtained from first
principles and one has to refer to other reactions involving 
the same coupling constants.
Such reactions are e.g. 
$\pi^{\pm} p \to a_2^{\pm} p$ (where both $I\!\!P \mp \rho^{0}$ exchanges are possible),
$\pi^{-} p \to a_{2}^{0} n$, $\pi^{-} p \to \omega^{0} n$ (only $\rho^{+}$-reggeon exchange come into game),
$\pi^{\pm} p \to \rho^{\pm} p$ ($\pi^{0}$, $\omega^{0}$- and $a_{2}^{0}$-reggeon exchanges)
and $\pi^{-} p \to \rho^{0} n$ ($\pi^{+}$, $a_{2}^{+}$-reggeon exchanges).
The details how to fix parameters of these two-body
reactions are described in the Appendix.


The diagram a) in Fig.\ref{fig:diagrams2} is topologically identical
to the dominant diagram for the  $p p \to  p p \pi^+ \pi^-$ reaction \cite{LS10}.
There, however, the pomeron-pomeron, pomeron-reggeon and reggeon-pomeron 
exchanges are the dominant processes.
In addition to diagram a) there is possible also another mechanism 
with the intermediate pion replaced by a virtual photon.
Because it requires two electromagnetic couplings instead of two strong
couplings its contribution should be small. Because of the extra photon 
propagator it could be enhanced when $k_{\gamma}^2 \to$ 0. However then 
the vertices should tend to zero. Therefore we can safely omit such 
a diagram.

\begin{figure}[!h]
a) \includegraphics[width = 0.18\textwidth]{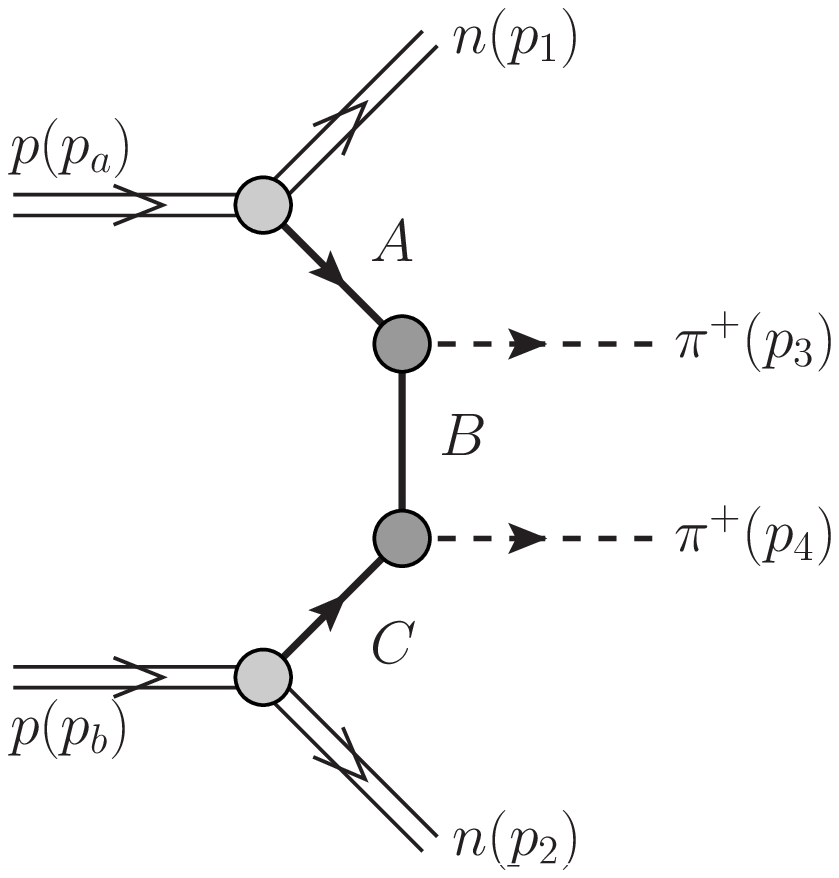}
b) \includegraphics[width = 0.18\textwidth]{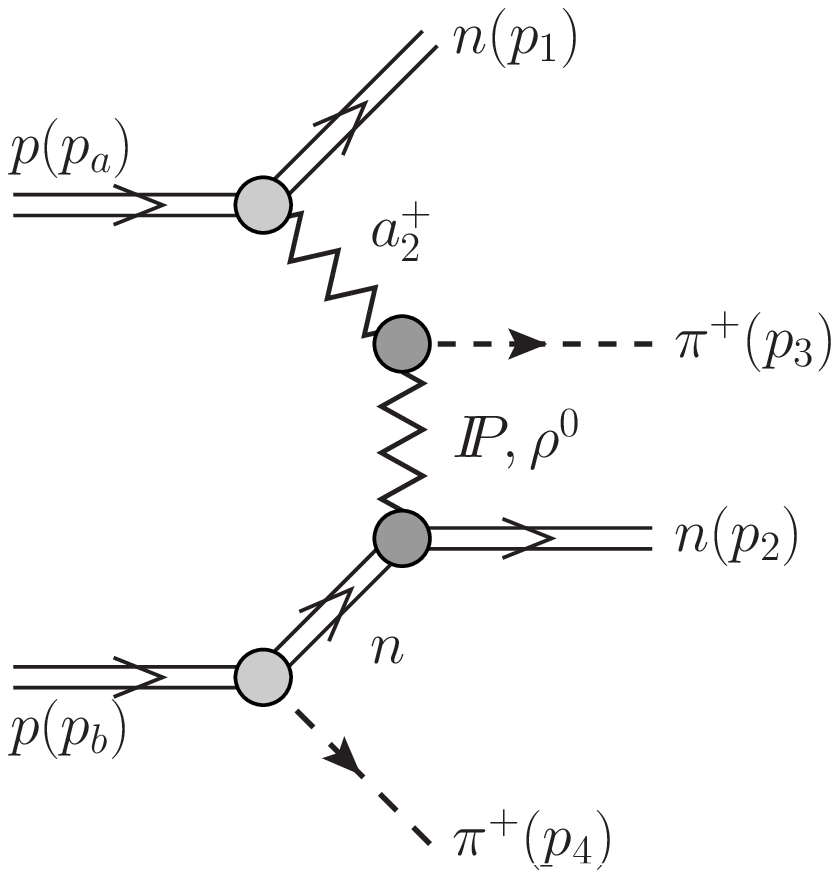}
c) \includegraphics[width = 0.18\textwidth]{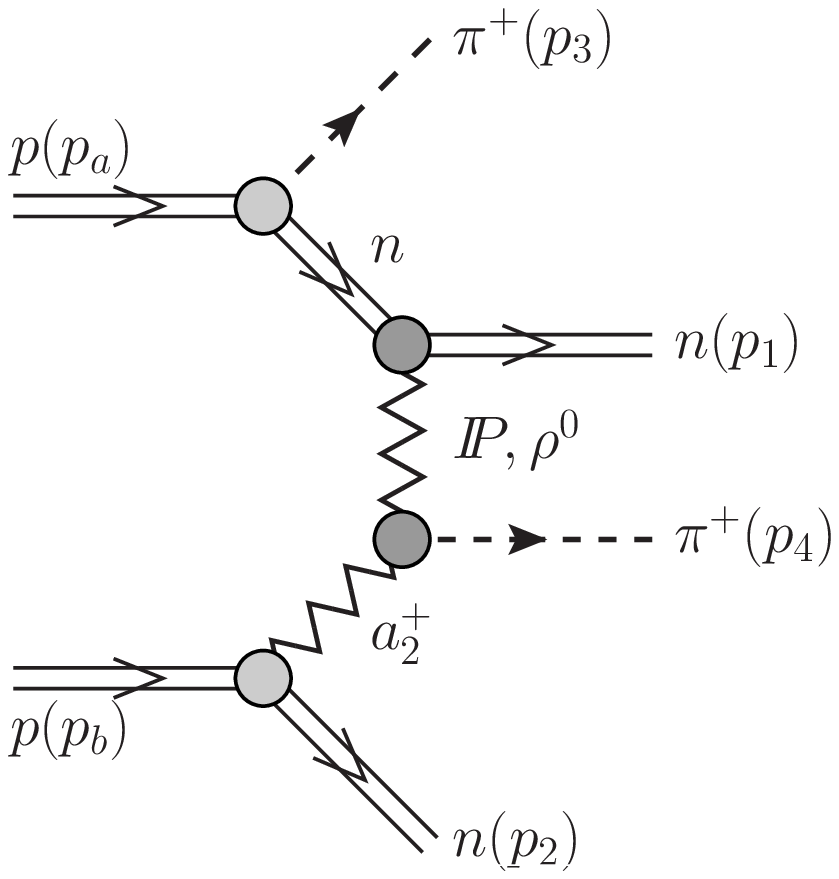}\\
d) \includegraphics[width = 0.22\textwidth]{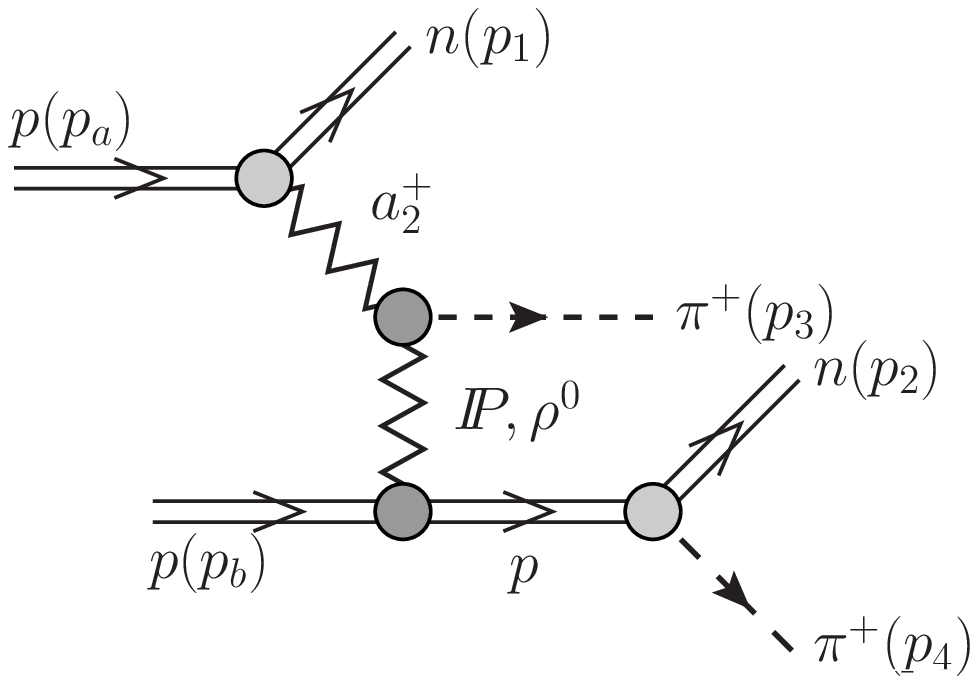}
e) \includegraphics[width = 0.22\textwidth]{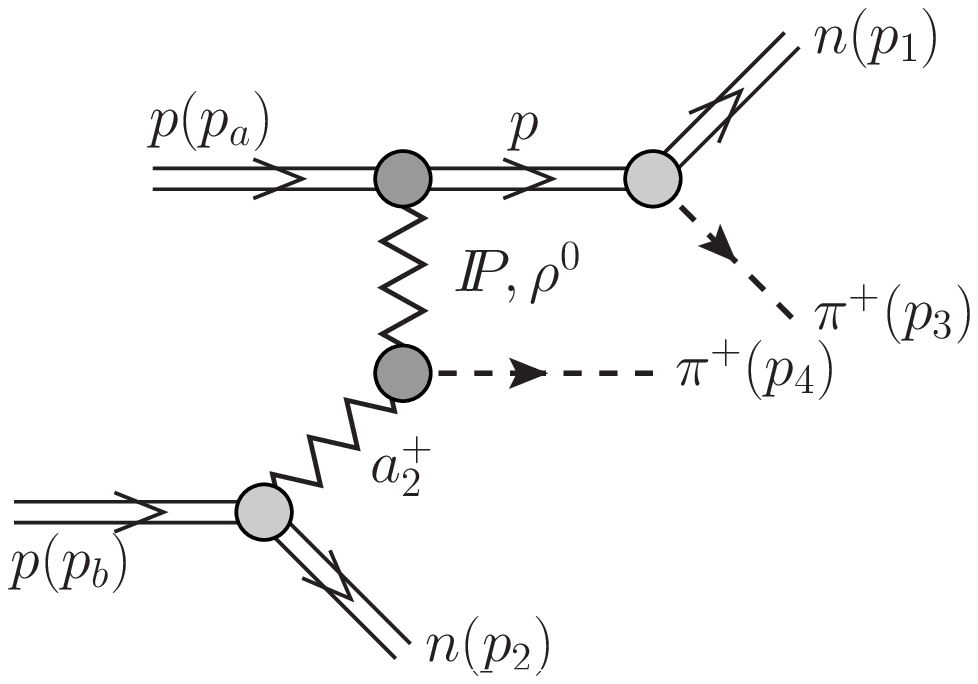}
  \caption{\label{fig:diagrams_other}
  \small
Diagrams with subleading reggeon $a_{2}^{+}$ exchange 
in $pp$ collisions at high energies.
}
\end{figure}

We write the amplitudes for the diagrams in Fig.\ref{fig:diagrams2} as:
\begin{eqnarray}
{\cal M}_{\lambda_{a}\lambda_{b} \to \lambda_{1}\lambda_{2}} &=&
\sqrt{2}  
\left(\frac{-t_{1}}{4 m_{N}^{2}}\right)^{|\lambda_{1}-\lambda_{a}|/2} 
r_{T}^{i\,|\lambda_{1}-\lambda_{a}|}\,
\eta_{I\!\!R}\, s_{13}\, \sqrt{C_{I\!\!R}^{NN}}
\left(\frac{s_{13}}{s_0}\right)^{\alpha_{I\!\!R}(t_{1})-1}
\exp \left( {\frac{B_{M N}}{2} t_{1}}\right) \nonumber \\
&\times & {\cal A}(s_{34},t_{a})\nonumber \\
&\times &
\sqrt{2}  
\left(\frac{-t_{2}}{4 m_{N}^{2}}\right)^{|\lambda_{2}-\lambda_{b}|/2} 
r_{T}^{i\,|\lambda_{2}-\lambda_{b}|}\,
\eta_{I\!\!R}\, s_{24}\, \sqrt{C_{I\!\!R}^{NN}}
\left(\frac{s_{24}}{s_0}\right)^{\alpha_{I\!\!R}(t_{2})-1}
\exp \left( {\frac{B_{M N}}{2} t_{2}}\right) \nonumber \\ 
&+ & crossed\,term\,,
\label{amp_a2a2}
\end{eqnarray} 
where ${\cal A}(s_{34},t_{a})$ refers to the central part of the diagrams
\begin{eqnarray}
{\cal A}^{\pi - exch.}(s_{34},t_{a}) &=&
F_{\pi}^{off}(t_{a}) \,
\sqrt{C_{\rho}^{\pi \pi}} \, 
\dfrac{1}{t_{a}-m_{\pi}^{2}} \,
 \sqrt{C_{\rho}^{\pi \pi}} \,
F_{\pi}^{off}(t_{a}) \,, 
\label{amp_pi}
\end{eqnarray}
\begin{eqnarray}
{\cal A}^{reggeon - exch.}(s_{34},t_{a}) &=&
 \frac{\sqrt{ -t_{a} }}{M_0} \,
 \eta_{i} \, s_{34} \, (g_{j \to \pi}^{i})^2 
\left(\frac{s_{34}}{s_0}\right)^{\alpha_{i}(t_{a})-1} 
 \exp \left( {\frac{B_{MM}}{2}  t_{a}  } \right) 
\frac{\sqrt{  -t_{a} }}{M_0} \,. 
\label{amp_rho}
\end{eqnarray}
In actual calculations we take $B_{MN} = B_{\pi N}$ and $B_{MM} = B_{\pi \pi}$.
Since, in the diagrams in
Fig.\ref{fig:diagrams2} and Fig.\ref{fig:diagrams_other} 
we have reggeon exchanges
rather than meson exchanges therefore
formulas (\ref{amp_a2a2}, \ref{amp_rho}) give rather upper limit for the cross section.

The parameterization of the amplitudes with subleading charged reggeon exchanges
cannot be used in the region of resonances in $\pi N$ or/and $\pi \pi$ subsystems \cite{LS10}.
Therefore, the amplitude used in the calculations must
contain restrictions on the four-body phase space.
To exclude the regions of resonances we modify the parameterization of the amplitudes 
(\ref{amp_a2a2})
by multiplying cross section by a purely phenomenological smooth cut-off correction factor (see \cite{LS10}):
\begin{eqnarray} 
f_{cont}^{\pi N/\pi \pi}(W_{\pi N/\pi \pi})=\frac{\exp \left( \frac{W-W_{0}}{a}\right)}{1+\exp \left( \frac{W-W_{0}}{a}\right)} \; .
\label{Correction_factor}
\end{eqnarray}
The parameter $W_{0}$ gives the position of the cut and 
the parameter $a$ describes how sharp is the cut off.
For large energies
$f_{cont}^{\pi N/\pi \pi}(W_{\pi N/\pi \pi})\approx$ 1 and close to
kinematical threshold
$f_{cont}^{\pi N/\pi \pi}(W_{\pi N/\pi \pi})\approx$ 0.
In our calculation we take $W_{0}=2$ GeV
and $a=0.2$ GeV.

There is another class of diagrams shown in Fig.\ref{fig:diagrams_other}.
The diagram (a) represents a generic amplitude, with particle sets
(A, B, C) collected in Table \ref{tab:tab_exchange}.
In contrast to the diagrams shown in Fig.\ref{fig:diagrams2}
here both pions and subleading reggeons couple to nucleons.
We shall not present explicit formulae for the corresponding amplitudes here.
We shall show separate contributions of those processes in the Result Section.

\begin{table}
\caption{Different realizations of diagram a) in Fig.\ref{fig:diagrams_other}.}
\label{tab:tab_exchange}
\begin{center}
\begin{tabular}{|c||c|c|c|c|c|}
\hline
A   & $a_{2}^{+}$ & $a_{2}^{+}$ & $\pi^{+}$  & $a_{2}^{+}$& $\pi^{+}$  \\
B   & $I\!\!P$    & $I\!\!P$    & $I\!\!P$   & $\rho^{0}$ & $\rho^{0}$ \\
C   & $a_{2}^{+}$ & $\pi^{+}$   & $a_{2}^{+}$& $\pi^{+}$  & $a_{2}^{+}$\\
\hline
\end{tabular}
\end{center}
\end{table}

\subsection{Absorptive corrections}
\label{subsection:absorptive_corrections}

\begin{figure}[!h]    %
\includegraphics[width=0.26\textwidth]{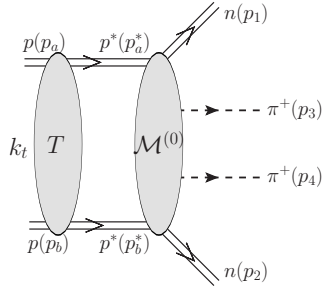}
  \caption{\label{fig:abs}
  \small
Schematic diagram for absorption effects due to proton-proton interaction.
}
\end{figure}

The absorptive correction in Fig.\ref{fig:abs} are calculated as described in \cite{SS07}
for the three body processes. 
Here the absorptive correction to the bare amplitude (see Fig.\ref{fig:diagrams})
can be writen as:
\begin{eqnarray}
{\delta\cal M}_{\lambda_{a}\lambda_{b} \to \lambda_{1}\lambda_{2}}(\vec{p}_{1t},\vec{p}_{2t} )=
\mathrm{i}
\int \frac{d^{2}k_{t}}{8 \pi^{2}} \frac{T(s, k_{t}^{2})}{s}
{\cal M}^{(0)}_{\lambda_{a}\lambda_{b} \to \lambda_{1}\lambda_{2}}(\vec{p}^{\,*}_{at}-\vec{p}_{1t},\vec{p}^{\,*}_{bt}-\vec{p}_{2t}) \;,
\label{abs_correction}
\end{eqnarray}
where $p^{\,*}_{a} = p_{a} - k_{t}$, $p^{\,*}_{b} = p_{b} + k_{t}$ with momentum transfer $k_{t}$.
Above ${\cal M}^{(0)}_{\lambda_{a}\lambda_{b} \to \lambda_{1}\lambda_{2}}$
is a bare amplitude calculated as described in the previous subsections.
$T(s, k_{t}^{2})$ is an elastic proton-proton amplitude
for the appropriate energy.
It can be conveniently parametrized as:
\begin{eqnarray}
T(s, k_{t}^{2})=
A_{0}(s)\;
\exp(-B_{NN} k_{t}^{2}/2)\;.
\label{T_factor}
\end{eqnarray}
From the optical theorem we have Im$A_{0}(s) = s \sigma^{pp}_{tot}(s)$
(the real part is small in the high energy limit).
Again the Donnachie-Landshoff parametrization \cite{DL92}
of the total $pp$ or $p\bar{p}$ cross sections
can be used to calculate the rescattering amplitude.

In our analysis the $\pi^{+} n$ interactions are not 
taken into account. They would further decrease
the cross section. Given other theoretical uncertainties (form factors)
it seems not worthy to take over the effort of performing very time-consuming calculations.
Absorption effects for exclusive Higgs production
are discussed e.g. in Ref.\cite{KMR_absorption}.

The cross section is obtained by assuming a general $2 \to 4$ reaction:
\begin{eqnarray}
\sigma =\int \frac{1}{2s} \overline{ |{\cal M}|^2} (2 \pi)^4
\delta^4 (p_a + p_b - p_1 - p_2 - p_3 - p_4)
\frac{d^3 p_1}{(2 \pi)^3 2 E_1}
\frac{d^3 p_2}{(2 \pi)^3 2 E_2}
\frac{d^3 p_3}{(2 \pi)^3 2 E_3}
\frac{d^3 p_4}{(2 \pi)^3 2 E_4}. \nonumber \\
\;
\label{dsigma_for_2to4}
\end{eqnarray}
To calculate the total cross section one has to calculate
8-dimensional integral numerically.
The details how to conveniently reduce the number
of kinematical integration variables are given elsewhere \cite{LS10}.
\section{Results}

We shall show our predictions for
the $pp \to nn \pi^+ \pi^+$ reaction
for several differential distributions in different
variables at selected center-of-mass energies
$W$ = 500 GeV (RHIC) and
$W$ = 0.9, 2.36 and 7 TeV (LHC).
The cross section slowly rises with incident energy.
In general, the higher energy the higher absorption effects.
The results depend on the value of the nonperturbative,
a priori unknown parameter of the form factor responsible for
off-shell effects.
In Table \ref{tab:sig_tot_ff} we have collected integrated cross sections
for selected energies and different values of the model parameters.
We show how the uncertainties of the form factor parameters
affect our final results.

\begin{table}
\caption{Full-phase-space integrated cross section (in mb)
for exclusive $nn \pi^+ \pi^+$ production at selected
center-of-mass energies
and different values of the form factor parameters.
In parentheses we show cross sections including absorption effects.}
\label{tab:sig_tot_ff}
\begin{center}
\begin{tabular}{|c||c|c|c|c|c|}
\hline
&W = 0.5 TeV &W = 0.9 TeV & W = 2.36 TeV & W = 7 TeV\\
\hline 
$\Lambda=0.8$ GeV, $\Lambda_{off}=1$ GeV & 0.34 (0.15) & 0.38 (0.16) & 0.47 (0.18) & 0.59 (0.19) \\
$\Lambda = \Lambda_{off}=1$ GeV          & 0.84 (0.37) & 0.95 (0.39) & 1.16 (0.42) & 1.47 (0.46) \\  
$\Lambda=1.2$ GeV, $\Lambda_{off}=1$ GeV & 1.45 (0.62) & 1.64 (0.66) & 2.01 (0.71) & 2.55 (0.77) \\
\hline
\end{tabular}
\end{center}
\end{table}

\begin{figure}[!h]
\includegraphics[width = 0.4\textwidth]{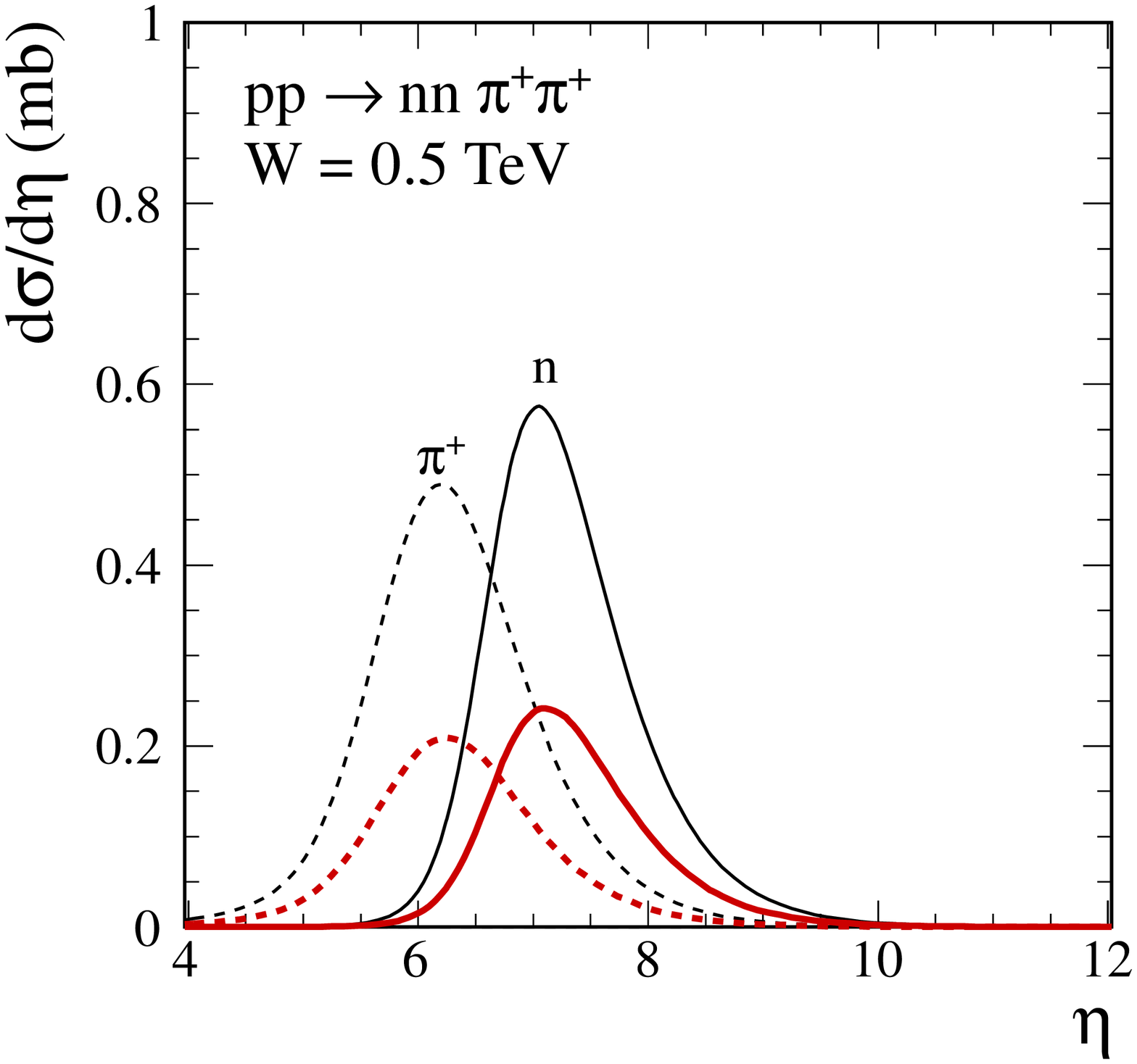}
\includegraphics[width = 0.4\textwidth]{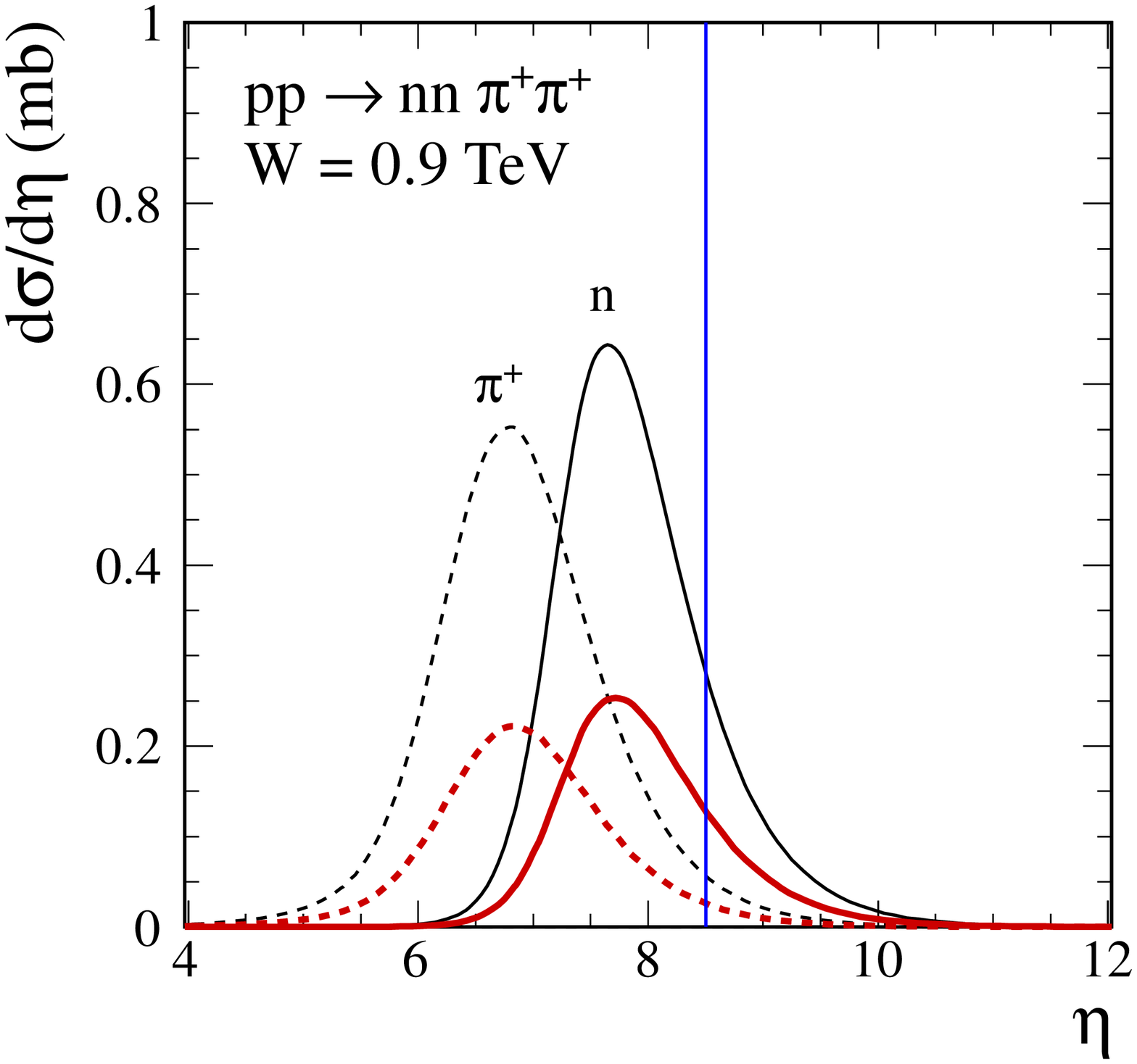}\\
\includegraphics[width = 0.4\textwidth]{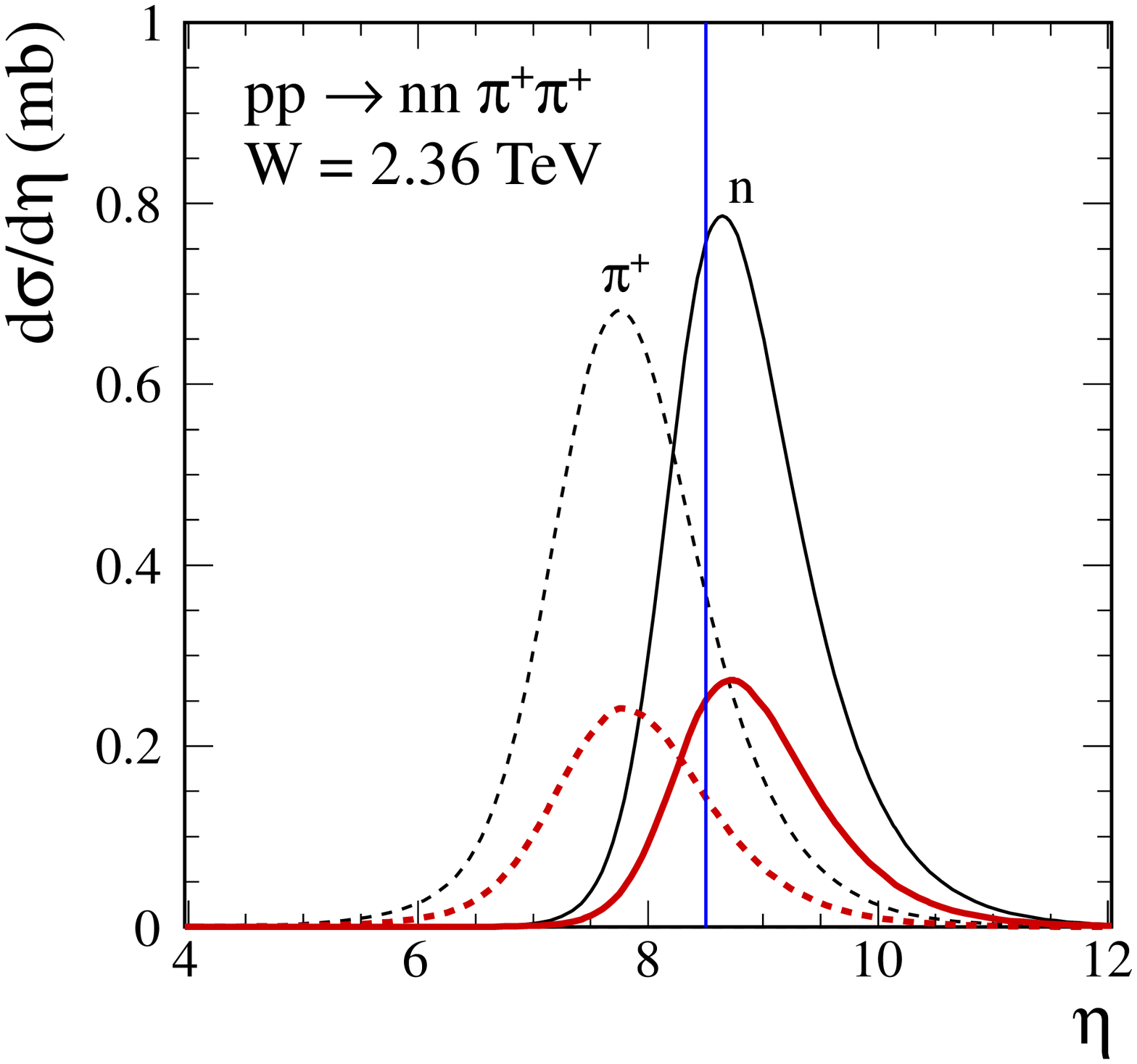}
\includegraphics[width = 0.4\textwidth]{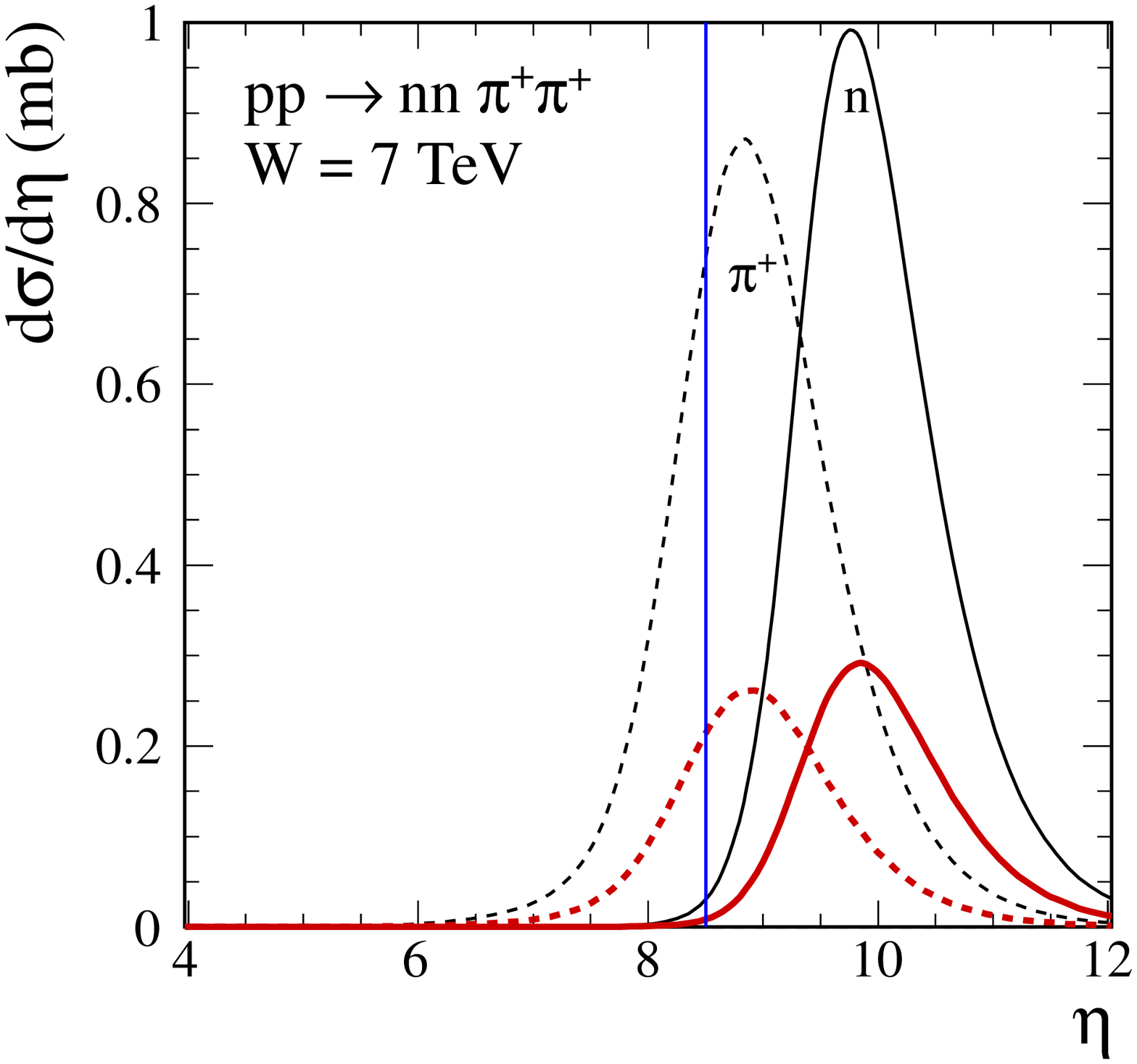}
  \caption{\label{fig:pseudorapidity_7000}
  \small
Differential cross section $d\sigma/d\eta$ for neutrons (solid lines)
and pions (dotted lines) at the center-of-mass energies $W$ = 0.5, 0.9, 2.36, 7 TeV.
The smaller bumps include absorption effects calculated in 
a way described in subsection \ref{subsection:absorptive_corrections}.
In this calculation we have used $\Lambda = \Lambda_{off}$ = 1 GeV.
The vertical lines at $\eta = \pm 8.5$ are the lower limits of the CMS ZDC's.
The details about RHIC ZDC's can be found in Ref.\protect\cite{ADGMSW01}.
}
\end{figure}

In Fig.\ref{fig:pseudorapidity_7000} we show distributions
in pseudorapidity
($\eta = -\ln (\tan \frac{\theta}{2})$,
where $\theta$ is the angle between the particle momentum and the beam axis)
for the $pp \to nn \pi^{+}\pi^{+}$ reaction.
The discussed reaction is very unique because not only
neutrons but also pions are produced dominantly in very forward or
very background directions forming a large size gap
in pseudorapidity between the produced pions, about 12
units at $W=7$ TeV.
While neutrons can be measured by the ZDC's the measurement
of very forward/backward pions requires further studies.
A possible evidence of the reaction discussed here is a signal from
both ZDC's and no signal in the central detector.

\begin{figure}[!h]
\includegraphics[width = 0.4\textwidth]{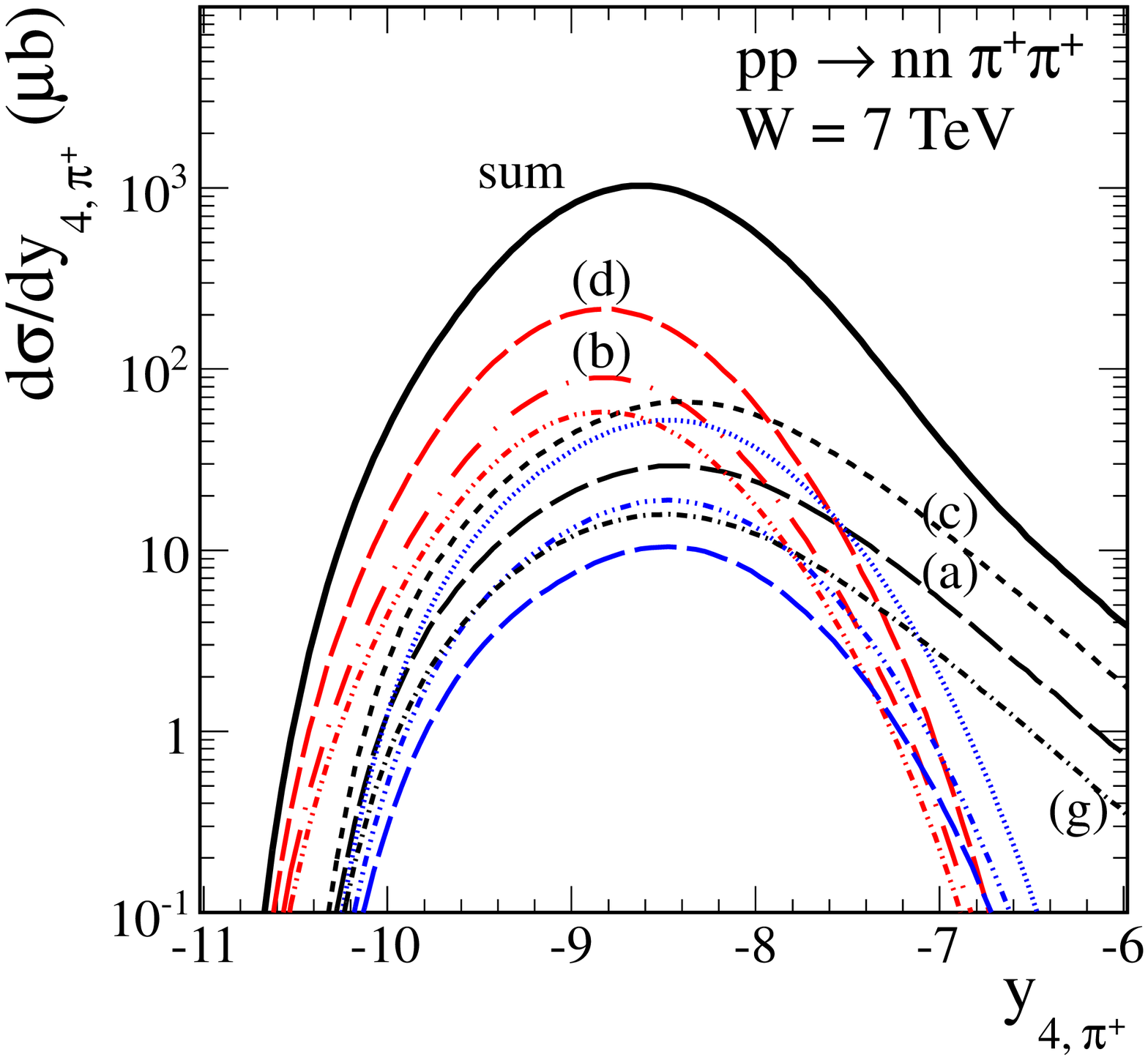}
\includegraphics[width = 0.4\textwidth]{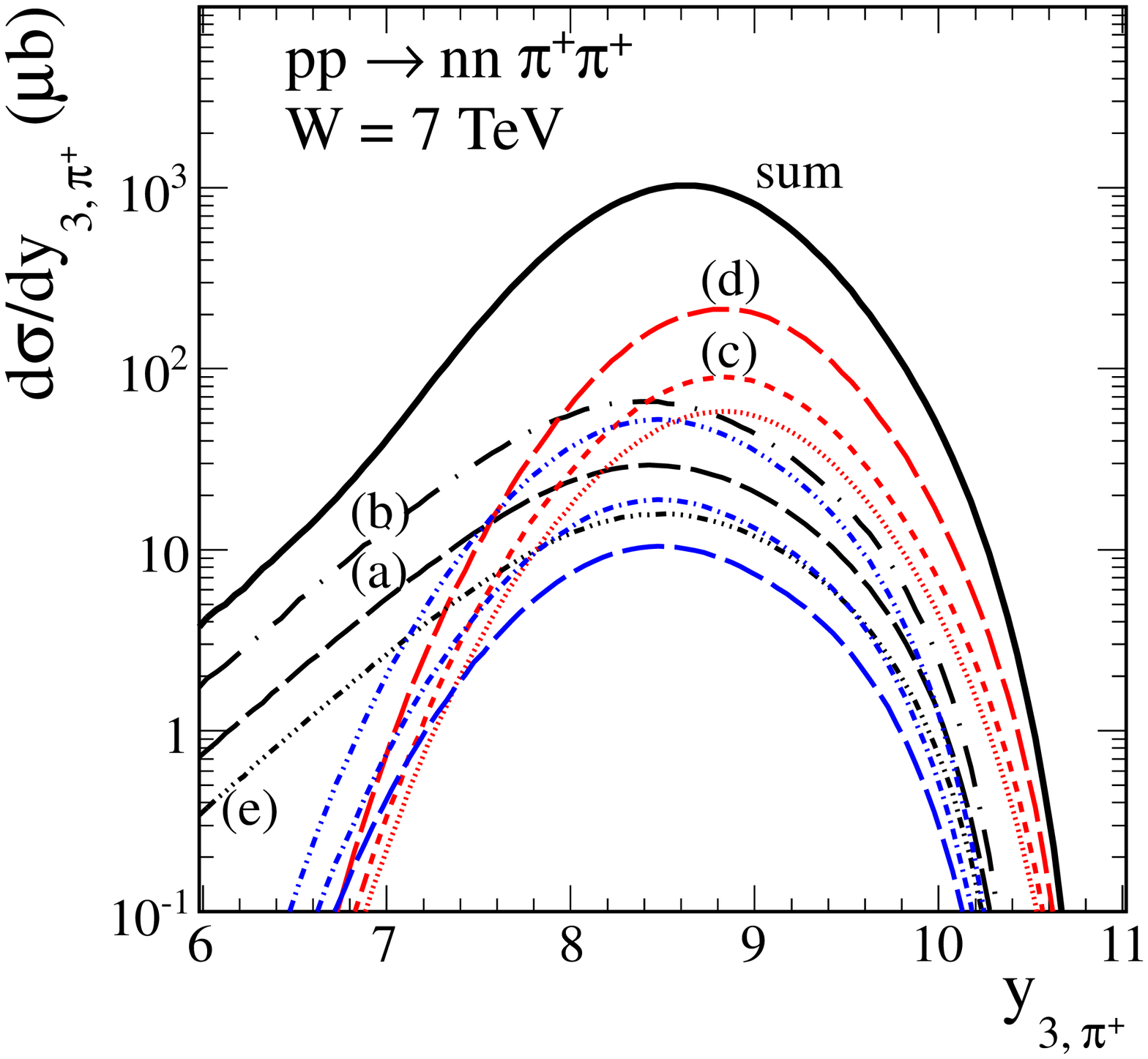}\\
\includegraphics[width = 0.4\textwidth]{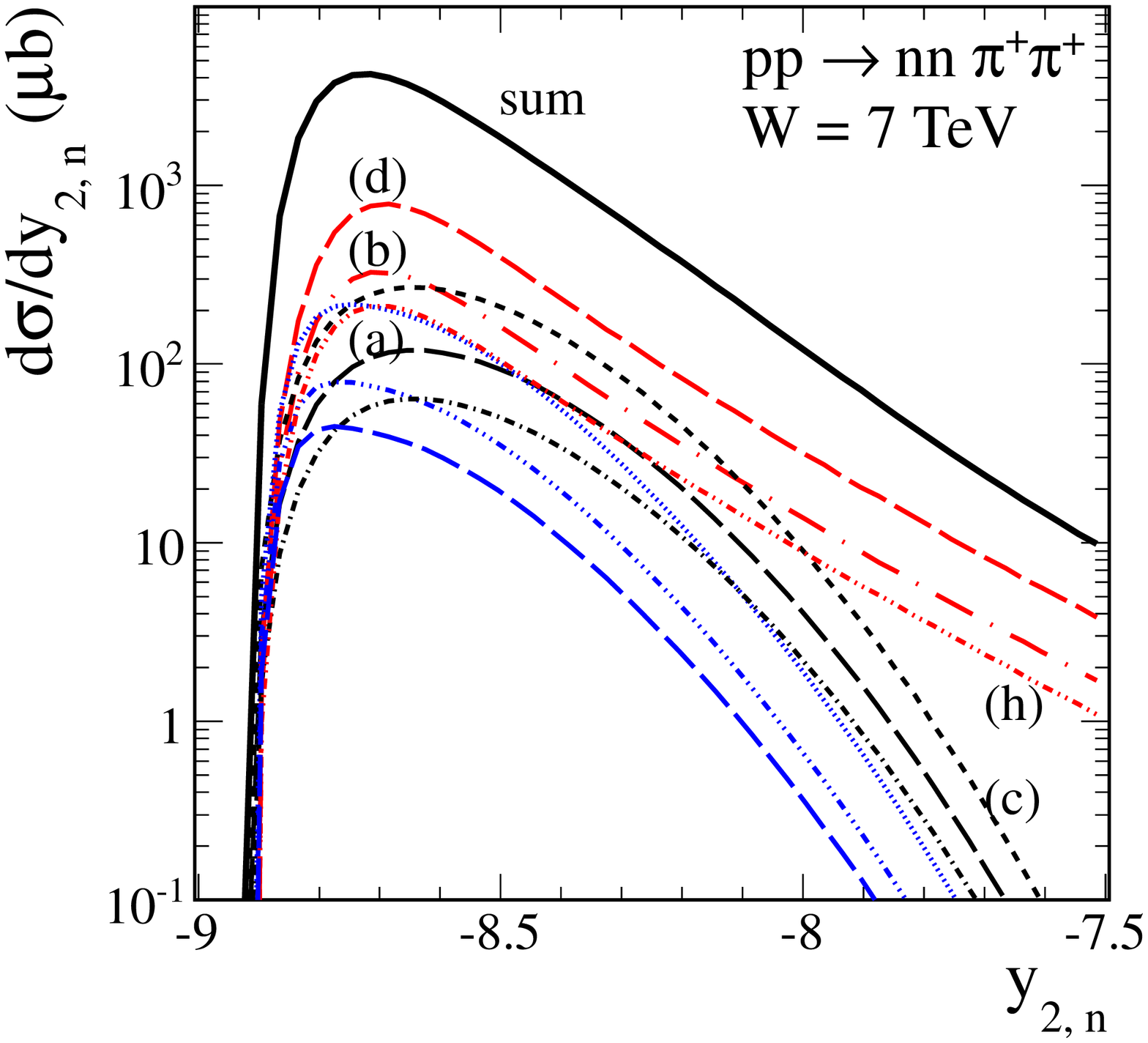}
\includegraphics[width = 0.4\textwidth]{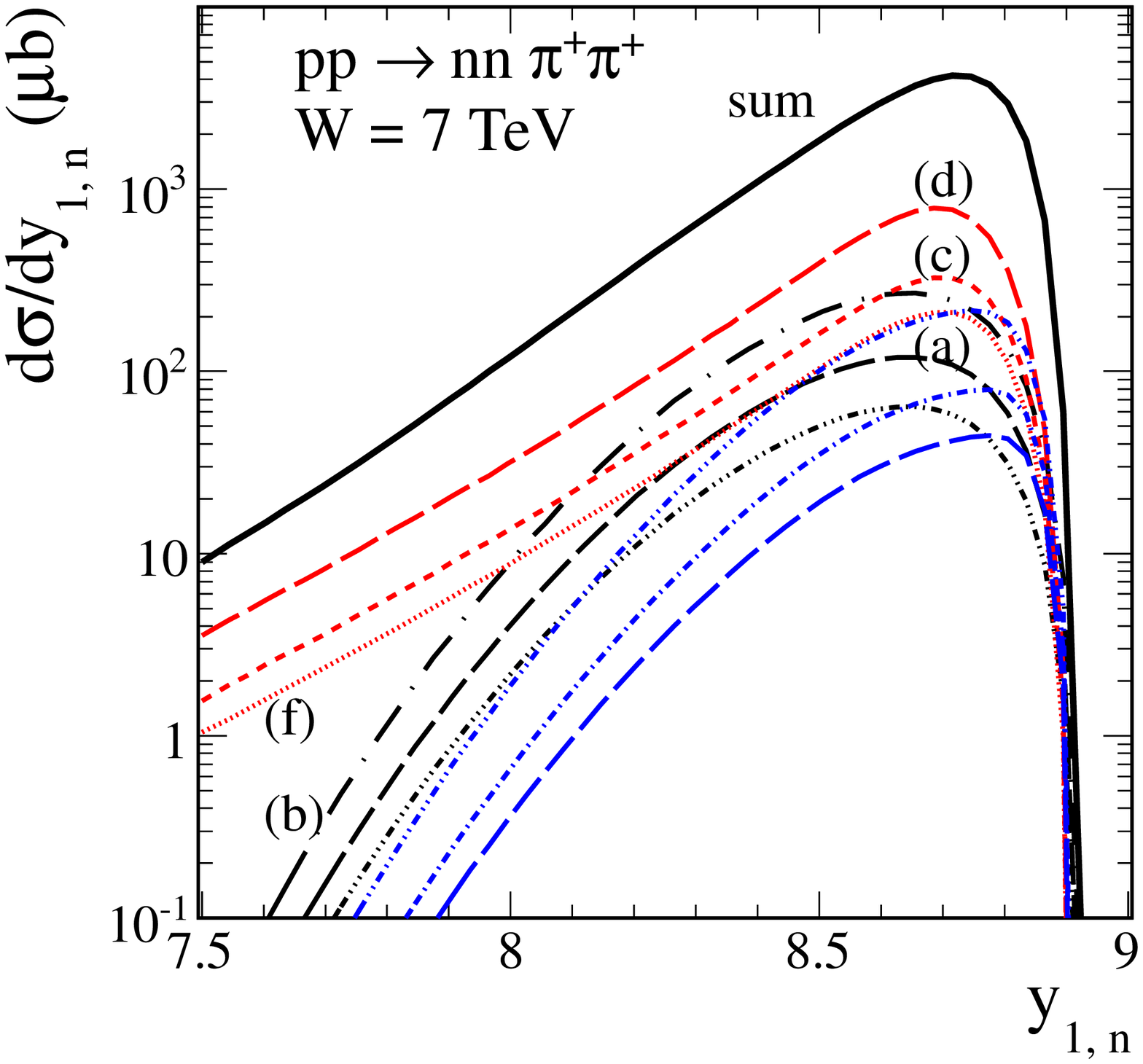}
  \caption{\label{fig:dsig_dy}
  \small
Differential cross sections $d\sigma/dy_{\pi^{+}}$ and $d\sigma/dy_{n}$ at $W$ = 7 TeV.
The bold solid line represent the coherent sum of all amplitudes.
The long-dashed (black), long-dash-dotted, dashed, long-dashed (red online),
dash-dot-dot-doted, dotted, dash-dotted, dash-dot-doted,
long-dashed (blue online) lines correspond to contributions from a) -- i) diagrams.
The red, black and blue lines correspond to diagrams when neutron,
pion and proton are off-mass-shell, respectively.
No absorption effects were included here.
}
\end{figure}

In Fig.\ref{fig:dsig_dy} we present rapidity distributions of pions
$y_{\pi^{+}}$ and rapidity distributions of neutrons $y_{n}$.
Please note a very limited range of rapidities shown in the figure.
The contributions for individual diagrams a) -- i)
(see Fig.\ref{fig:diagrams}) are also shown.
The diagram d) (from Fig.\ref{fig:diagrams}) gives the largest contribution.
One can observe specific symmetries between
different contributions on the left and right panels.
For instance the long-dash-dotted line on the left
panel (corresponding to diagram b) ) is symmetric
to the dashed line on the right panel
(corresponding to diagram c) ).
Clearly, a significant interference effect can be seen.
There is no region of either pion or neutron
rapidity where the diagram (a) dominates.
This makes the possibility of extracting of $\pi^+ \pi^+$
elastic scattering very difficult.

\begin{figure}[!h]
\includegraphics[width = 0.4\textwidth]{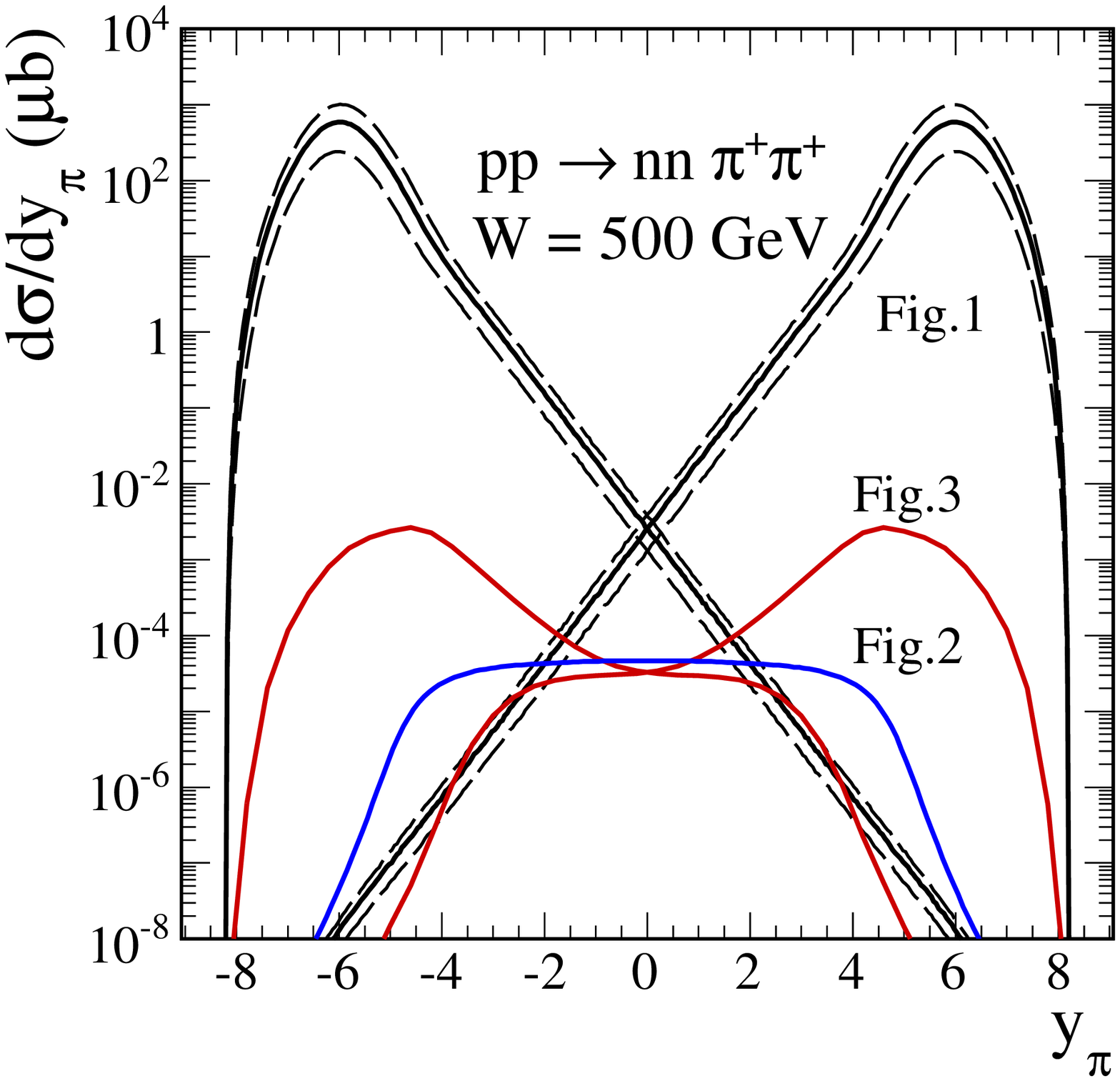}
\includegraphics[width = 0.4\textwidth]{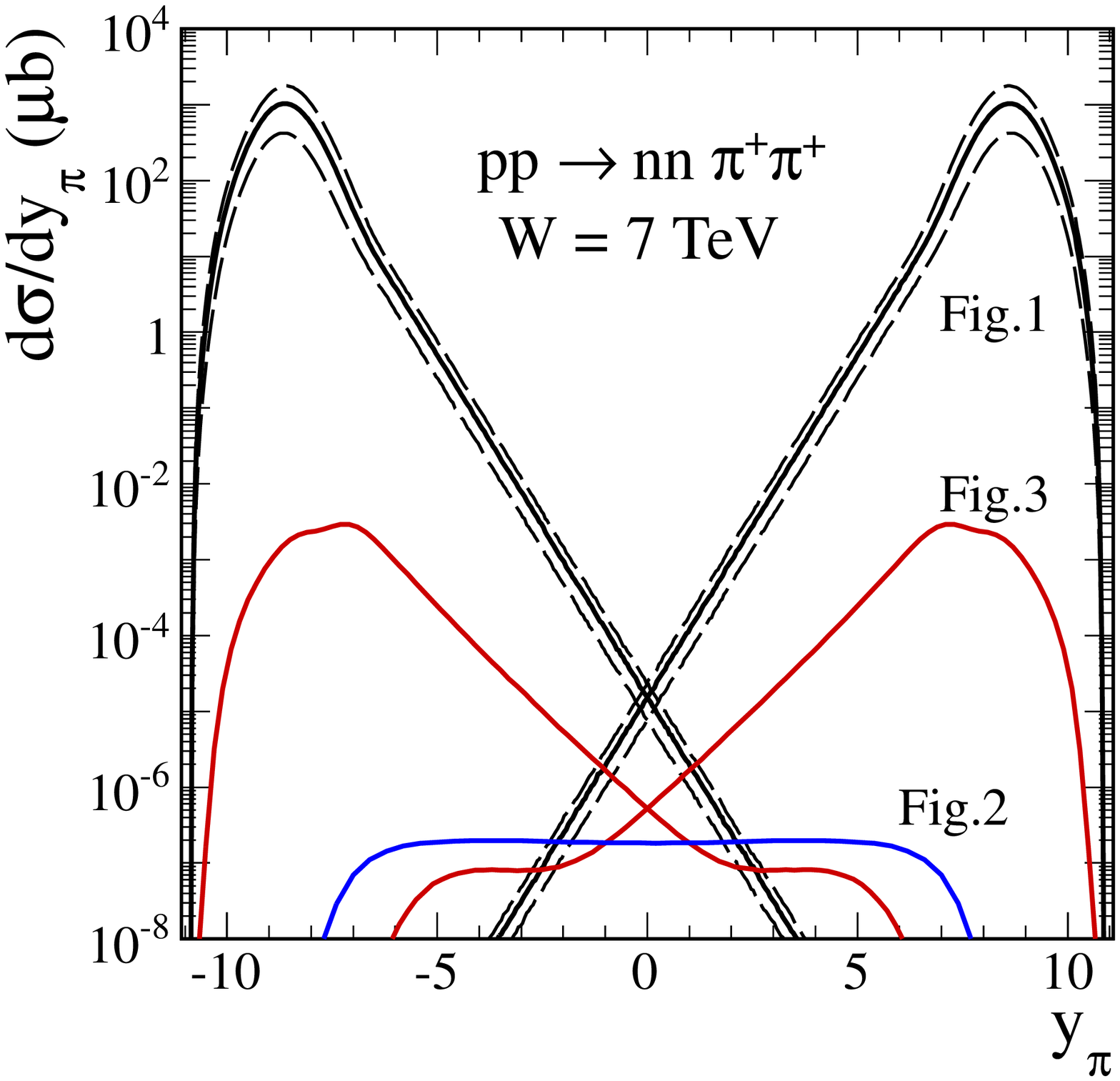}
  \caption{\label{fig:dsig_dy_pion}
  \small
Differential cross sections $d\sigma/dy_{\pi^{+}}$ 
at $W$ = 500 GeV (left) and $W$ = 7 TeV (right).
The lines represent the coherent sum of all amplitudes 
from diagrams in Fig.\ref{fig:diagrams}$-$\ref{fig:diagrams_other},
as well as separate classes of contributions marked by the number
of the figure where they are shown.
No absorption effects were included here.
}
\end{figure}

For completeness in Fig.\ref{fig:dsig_dy_pion} we show the contribution
of the diagrams with subleading charged reggeon exchanges (see Fig.\ref{fig:diagrams2})
which could not be seen in the previous plot.
We show results for the RHIC (left panel) and LHC (right panel) energies.
In contrast to the other mechanisms
the corresponding contribution is rather flat over broad range of rapidities. 
The cross section corresponding to this mechanism 
is bigger by 2 orders of magnitude for the RHIC energy compared to the LHC energy, 
but rather small compared to the dominant contributions shown in Fig.\ref{fig:diagrams}.
In addition we show contribution of diagrams of Fig.\ref{fig:diagrams_other}.
They are comparable to those of diagrams shown in Fig.\ref{fig:diagrams2}
at midrapidities but much smaller than those from Fig.\ref{fig:diagrams}
at larger rapidities.
We show results of diagrams from Fig.\ref{fig:diagrams} with different values 
of the form factor parameter $\Lambda = 0.8$ GeV (bottom dashed line) and 
$\Lambda = 1.2$ GeV (upper dashed line) in order to demonstrate the cross section uncertainties.

\begin{figure}[!h]
\includegraphics[width = 0.4\textwidth]{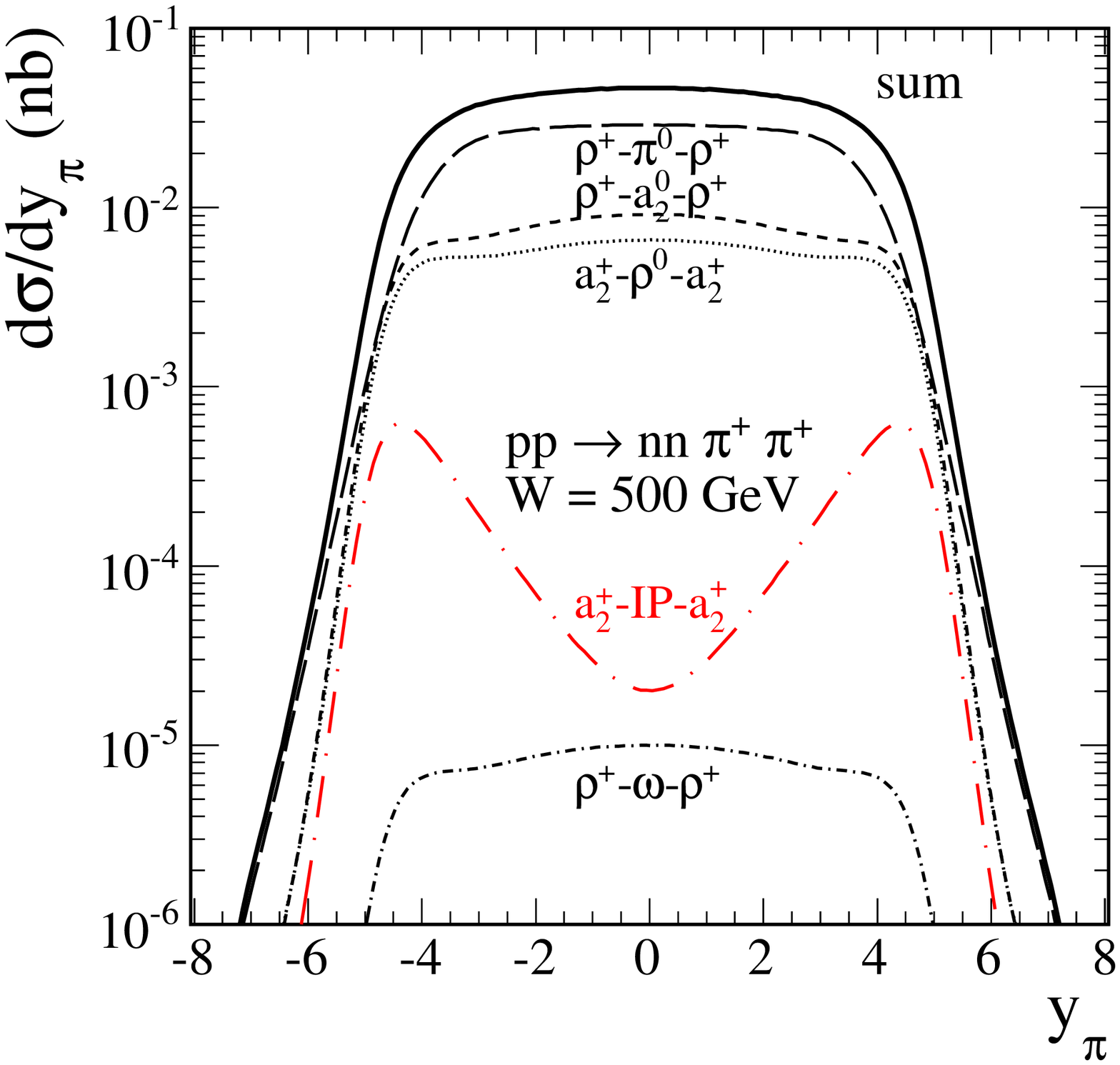}
\includegraphics[width = 0.4\textwidth]{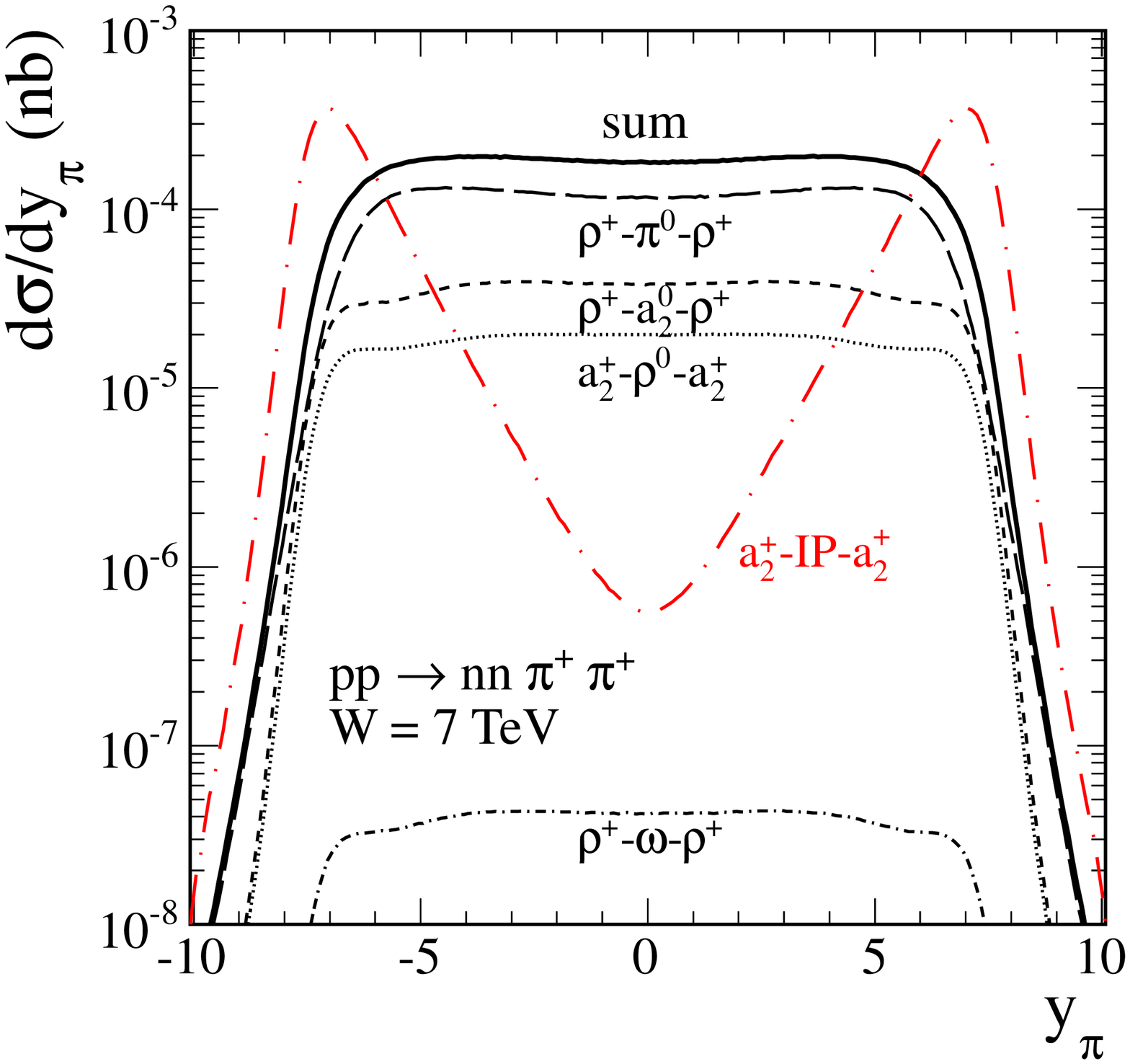}
  \caption{\label{fig:dsig_dy_charge_exch}
  \small
Differential cross sections $d\sigma/dy_{\pi}$
for double charged reggeon exchanges
at $W$ = 500 GeV (left) and $W$ = 7 TeV (right).
The contributions for individual diagrams
in Fig.\ref{fig:diagrams2} are shown separately.
No absorption effects were included here.
}
\end{figure}

In Fig.\ref{fig:dsig_dy_charge_exch} 
we present rapidity distributions of pions $y_{\pi}$
for double charged reggeon exchanges 
at $W$ = 500 GeV (left panel) and $W$ = 7 TeV (right panel).
The bold solid line represent the coherent sum of all amplitudes
corresponding to diagrams in Fig.\ref{fig:diagrams2}.
The contributions for individual diagrams 
are also shown separately.
The diagram a) in Fig.\ref{fig:diagrams2}
gives the largest contribution (long-dashed line).
The $a_{2}^{+} - I\!\!P - a_{2}^{+}$ exchange
corresponds to the long-dashed-dotted line.
One can see that the double reggeon exchange mechanisms
shown in Fig.\ref{fig:diagrams2} populate midrapidities
of the pions and therefore can be measured either at RHIC or at LHC. 
In Table \ref{tab:sig_tot_decomposition} we have
collected cross section for this component separately
for double spin conserving (DSC), single spin flip (SSF) and
double spin flip (DSF) contributions. All this spin contributions
are of similar size. The total contribution is about half of nb
at RHIC (500 GeV) and a few pb at LHC (7 TeV).
\begin{table}
\caption{Full-phase-space integrated cross section (in nb)
for exclusive $\pi^+ \pi^+$ production 
for the amplitude with the double charged reggeon exchanges (diagrams in Fig.\ref{fig:diagrams2})
at the center-of-mass energies $W$ = 0.5, 7 TeV.
No absorption effects were included here.
The meaning of the acronyms:
DSC - double spin conserving, SSF - single spin flip, DSF - double spin flip.}
\label{tab:sig_tot_decomposition}
\begin{center}
\begin{tabular}{|c||c|c|c|c|}
\hline
exchange & W = 0.5 TeV & W = 7 TeV \\
\hline
$\rho^{+}-\pi^{0}-\rho^{+}$      & 0.43             & 3.3$\times$10$^{-3}$\\
$\rho^{+}-a_{2}^{0}-\rho^{+}$    & 0.14             & 1.0$\times$10$^{-3}$\\
$a_{2}^{+}-\rho^{0}-a_{2}^{+}$   & 0.11             & 5.4$\times$10$^{-4}$\\
$\rho^{+}-\omega-\rho^{+}$   & 1.5$\times$10$^{-4}$ & 1.1$\times$10$^{-6}$\\
\hline
sum of all amplitudes & 0.7 & 5.1$\times$10$^{-3}$ \\
\hline
\hline
DSC  & 0.17 & 1.5$\times$10$^{-3}$\\
SSF  & 0.18 & 1.3$\times$10$^{-3}$\\  
DSF  & 0.18 & 1.0$\times$10$^{-3}$\\
\hline
\hline
$a_{2}^{+}-I\!\!P-a_{2}^{+}$ & 4.4$\times$10$^{-3}$ & 2.5$\times$10$^{-3}$\\
\hline
\end{tabular}
\end{center}
\end{table}

\begin{figure}[!h]
\includegraphics[width = 0.4\textwidth]{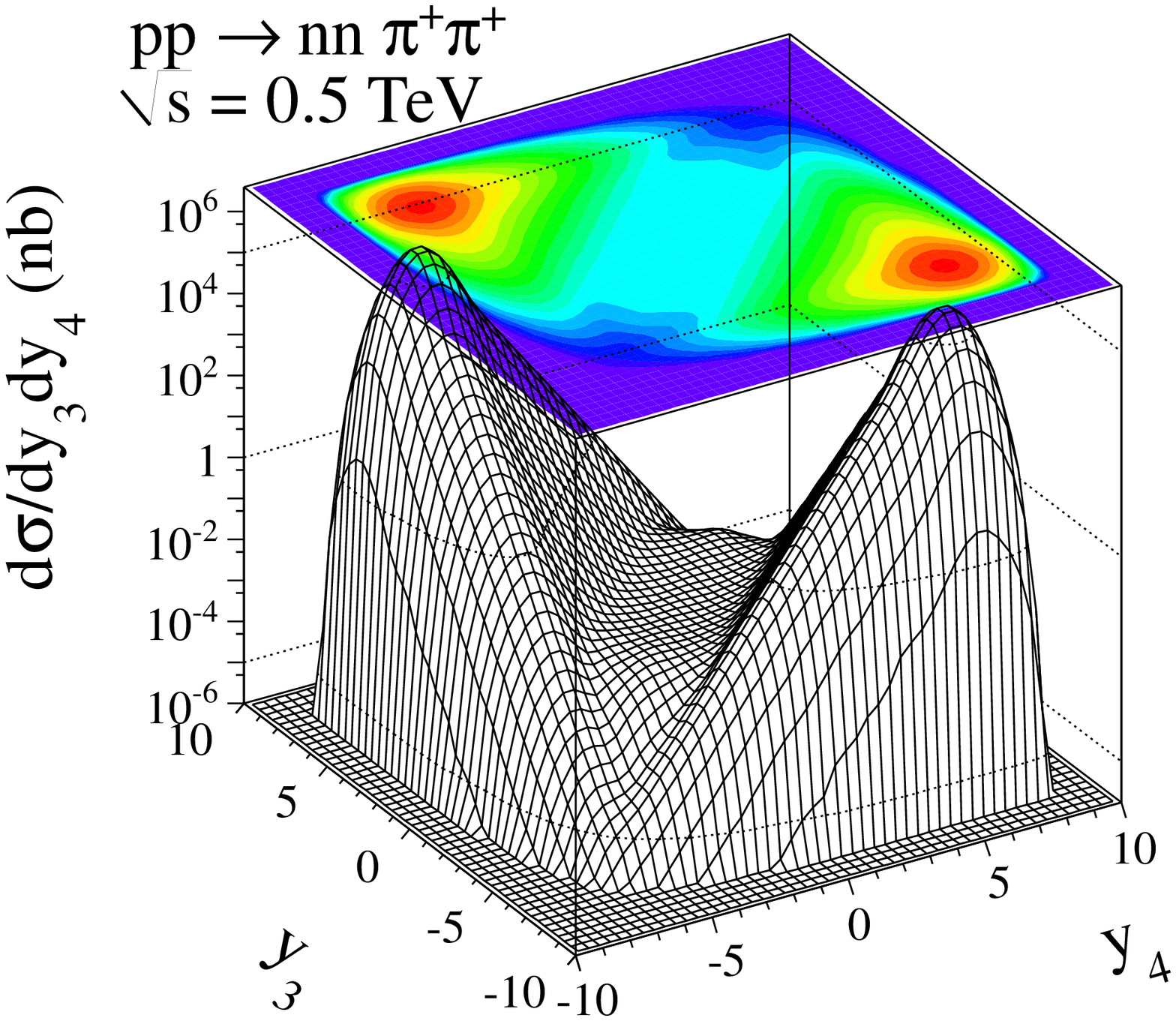}
\includegraphics[width = 0.4\textwidth]{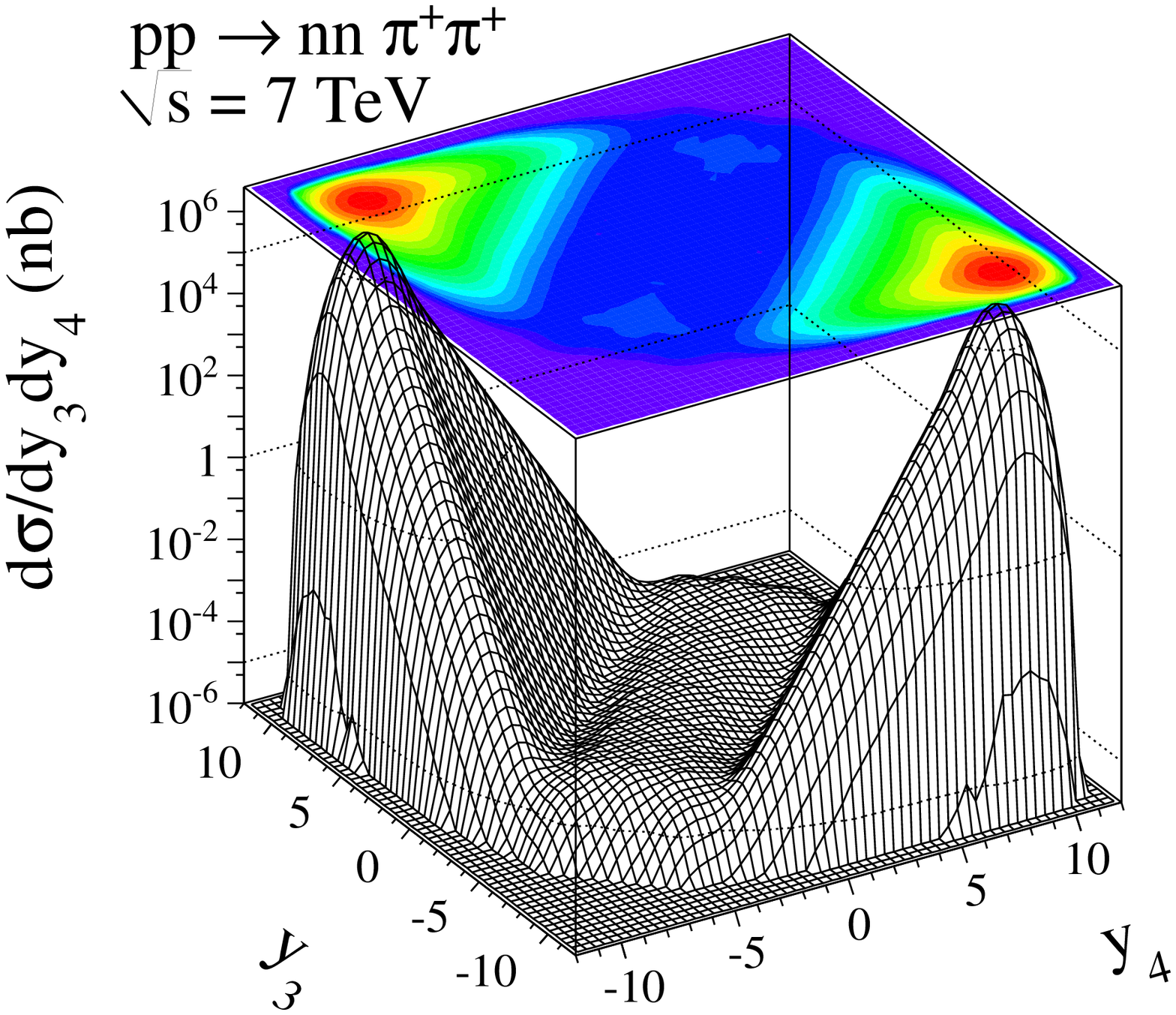}
  \caption{\label{fig:dsig_dy_2d_pion}
  \small
Differential cross sections in $(y_{3},y_{4})$ space 
at $W$ = 500 GeV (left) and $W$ = 7 TeV (right).
The coherent sum of all amplitudes from diagrams in Fig.\ref{fig:diagrams},
Fig.\ref{fig:diagrams_other}
and the contribution of diagrams in Fig.\ref{fig:diagrams2} with
double-exchange reggeons placed along the diagonal are presented.
No absorption effects were included here.
}
\end{figure}

\begin{figure}[!h]
\includegraphics[width = 0.4\textwidth]{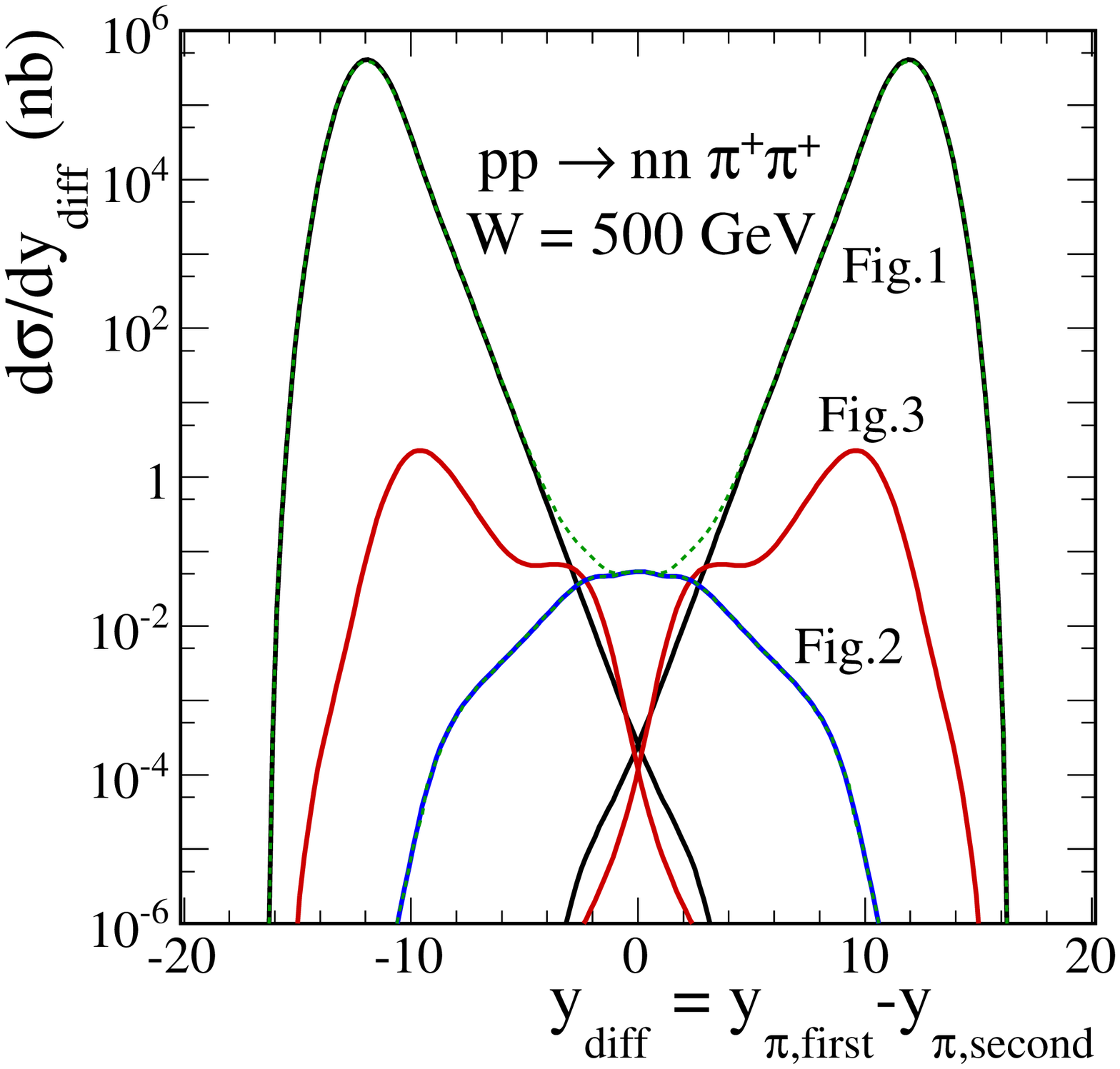}
\includegraphics[width = 0.4\textwidth]{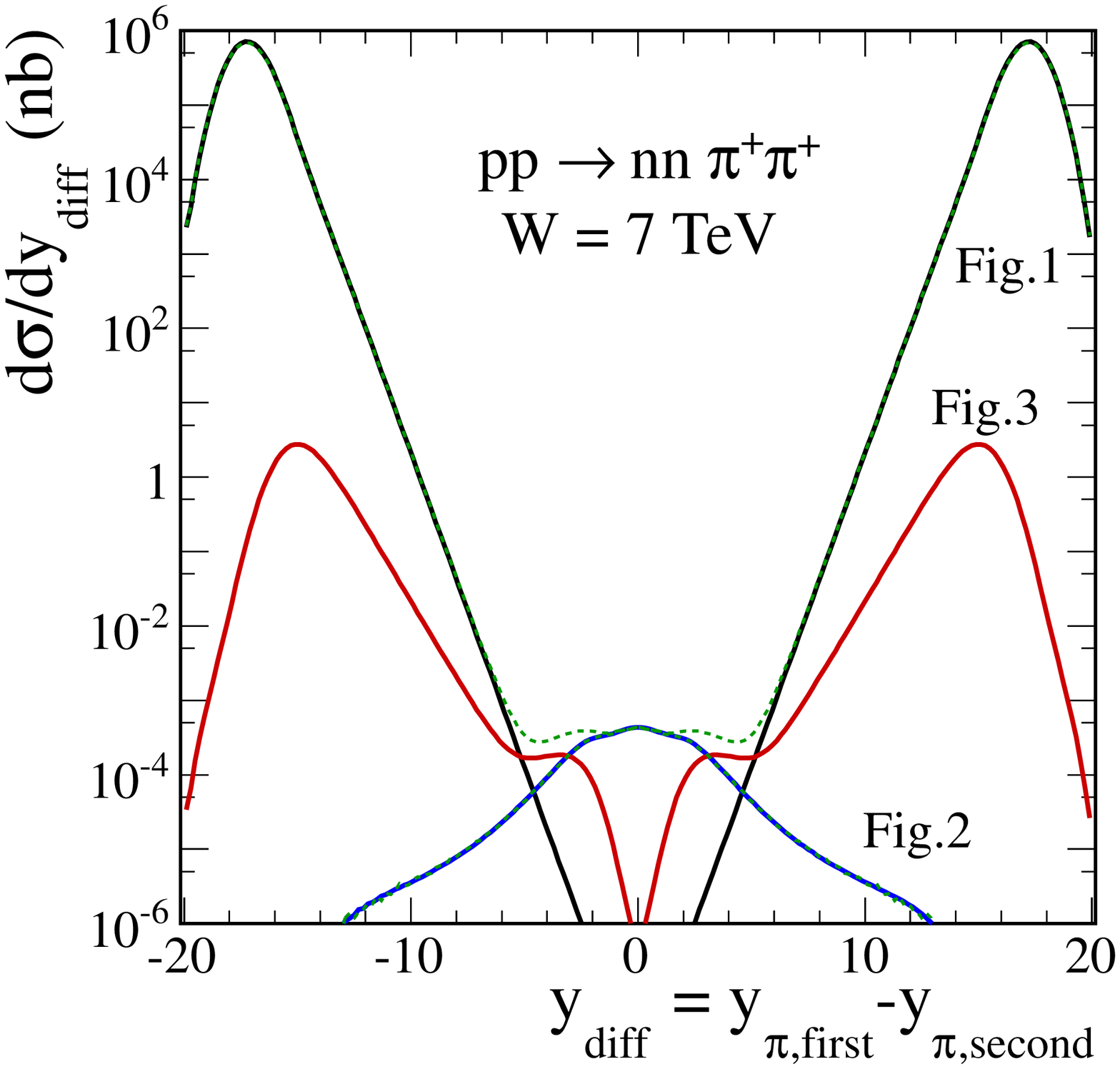}
  \caption{\label{fig:dsig_dydiff_pion}
  \small
Differential cross sections $d\sigma/dy_{diff}$ at $W$ = 500 GeV (left) and $W$ = 7 TeV (right).
The lines represent the coherent sum of all amplitudes 
from diagrams in Fig.\ref{fig:diagrams}, Fig.\ref{fig:diagrams_other}
and the contribution of diagrams in Fig.\ref{fig:diagrams2} with
double-exchange reggeons placed at $y_{diff} \approx 0$.
No absorption effects were included here.
}
\end{figure}

Can the much smaller contribution of diagrams 
with subleading charged reggeon exchanges be identified experimentally?
In Fig.\ref{fig:dsig_dy_2d_pion} we show two-dimensional
distribution in $(y_{3},y_{4})$ space.
The double-charged reggeon-exchange components from
Fig.\ref{fig:diagrams2} are placed along the diagonal
$y_{3} = y_{4}$ while the other contributions
some distance from the diagonal.
Therefore imposing 2-dim cuts in the $(y_{3},y_{4})$ space one could separate
the small double charged reggeons contribution.
A very good one-dimensional observable which can be 
used for the separation of the processes under disscusion could be
differential cross section $d\sigma/dy_{diff}$,
where $y_{diff}=y_{3}-y_{4}$ and experimentally charged pions should be taken at random (see Fig.\ref{fig:dsig_dydiff_pion},
$y_{\pi,first}=y_{3}$ or $y_{4}$ and $y_{\pi,second}=y_{4}$ or $y_{3}$).
For comparison we show contribution of diagrams shown in Fig.\ref{fig:diagrams_other}.

\begin{figure}[!h]
\includegraphics[width = 0.4\textwidth]{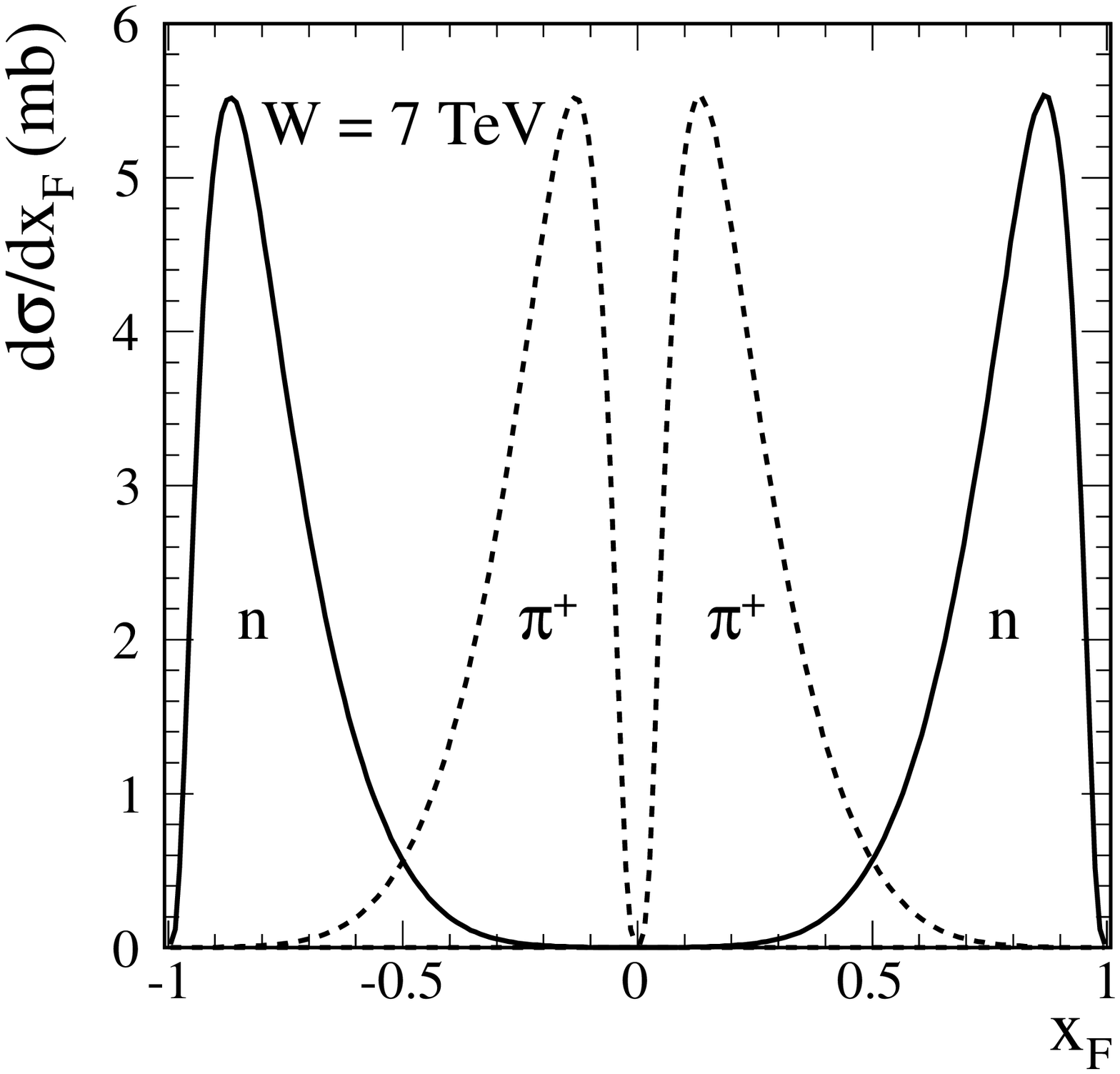}
  \caption{\label{fig:dsig_dxf}
  \small
Differential cross section $d\sigma/dx_{F}$
for the $pp \to nn \pi^{+}\pi^{+}$ reaction
at $W$ = 7 TeV.
No absorption effects were included here.
}
\end{figure}

In Fig.\ref{fig:dsig_dxf} we show distribution of neutrons and pions
in the Feynman variable $x_F = 2p_{\parallel}/\sqrt{s}$. 
In this observable the neutrons and pions are well separated. 
The position of peaks is almost independent of energy.
While pions are produce at relatively small $x_F$ the neutrons carry large fractions
of the parent protons. The situation is qualitatively the same for all
energies. 

\begin{figure}[!h]  
\includegraphics[width = 0.4\textwidth]{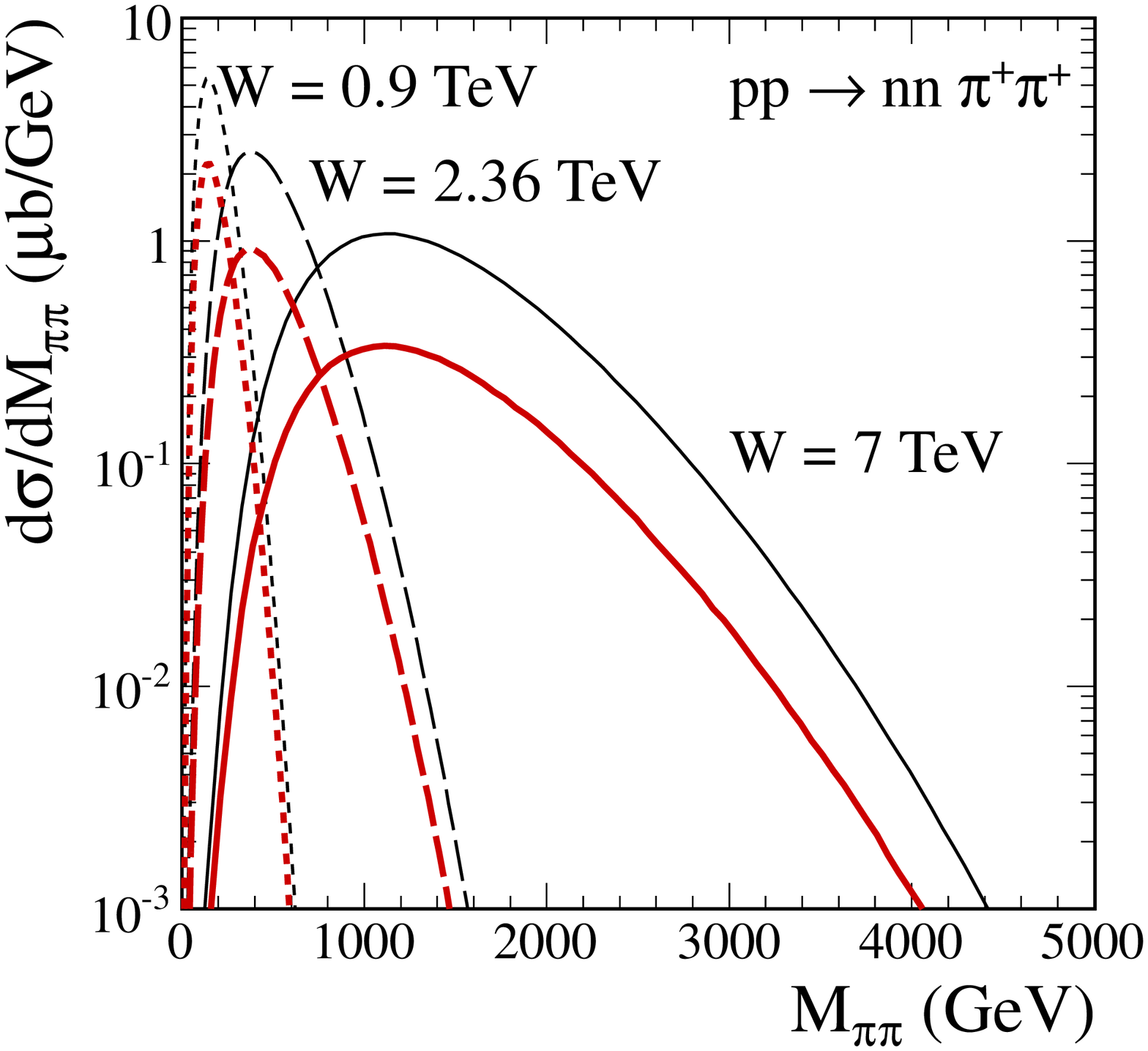}
\includegraphics[width = 0.4\textwidth]{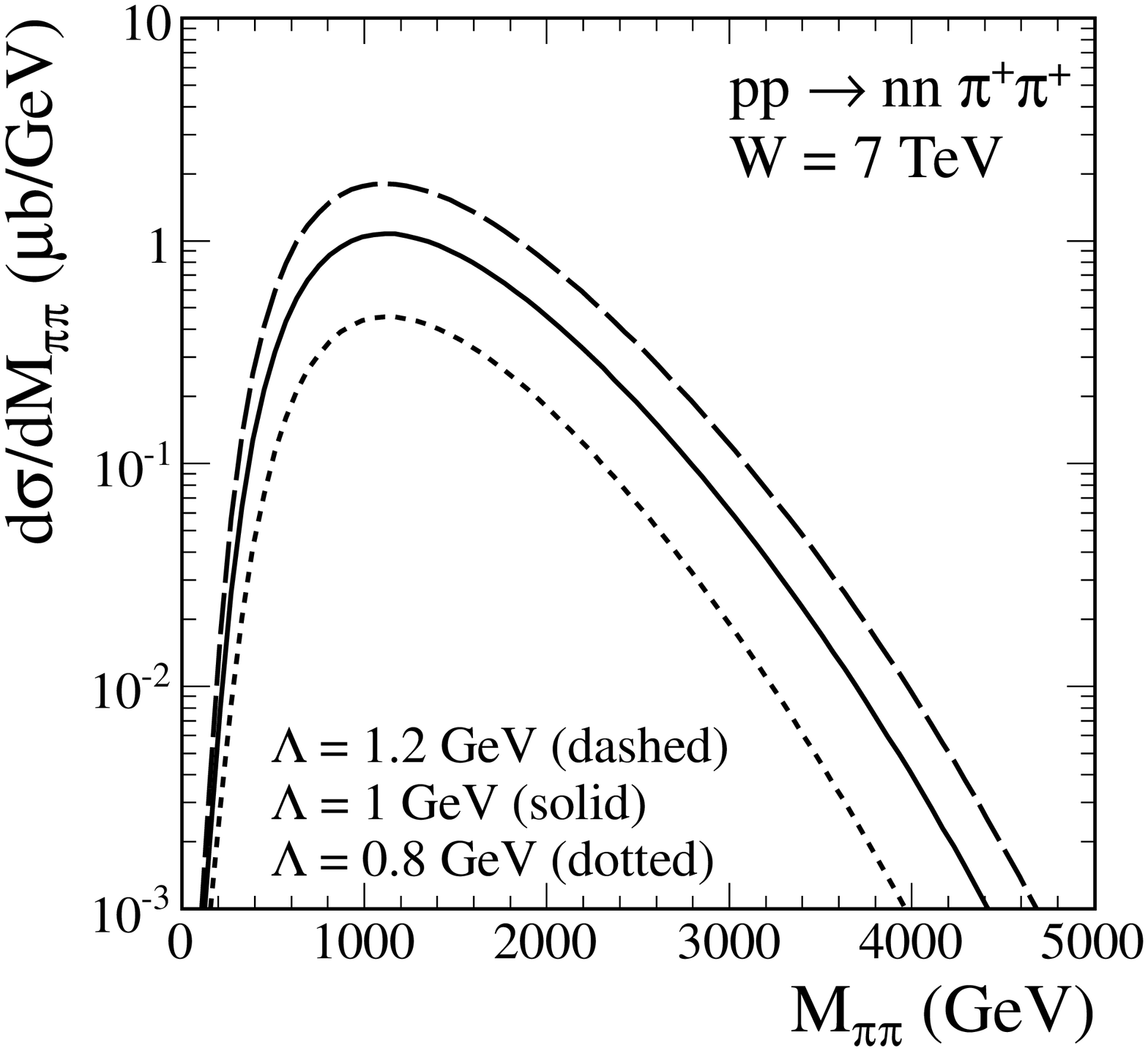}
  \caption{\label{fig:dsig_dm34}
  \small
Differential cross section $d\sigma/dM_{\pi\pi}$
for the $pp \to nn \pi^{+}\pi^{+}$ reaction
at $W$ = 0.9, 2.36, 7 TeV (left panel).
The lower curves correspond to calculations
with absorption effects.
Right panel shows "bare" cross section obtained with different values
of the form factor parameter $\Lambda = 0.8$ GeV (doted line),
$\Lambda = 1$ GeV (solid line) and $\Lambda = 1.2$ GeV (dashed line)
at $W$ = 7 TeV..
}
\end{figure}

The distribution in pion-pion invariant mass is shown in
Fig.\ref{fig:dsig_dm34}.
Unique for this reaction, very large two-pion
invariant masses are produced (see e.g. Ref.\cite{LS10}).
The larger energy the larger two-pion invariant masses
(left panel).
The absorption effects almost uniformly reduce
the cross section.
We show also distributions
with different values of the form factor parameter
in order to demonstrate the cross section uncertainties (righ panel).

\begin{figure}[!h]  
\includegraphics[width = 0.4\textwidth]{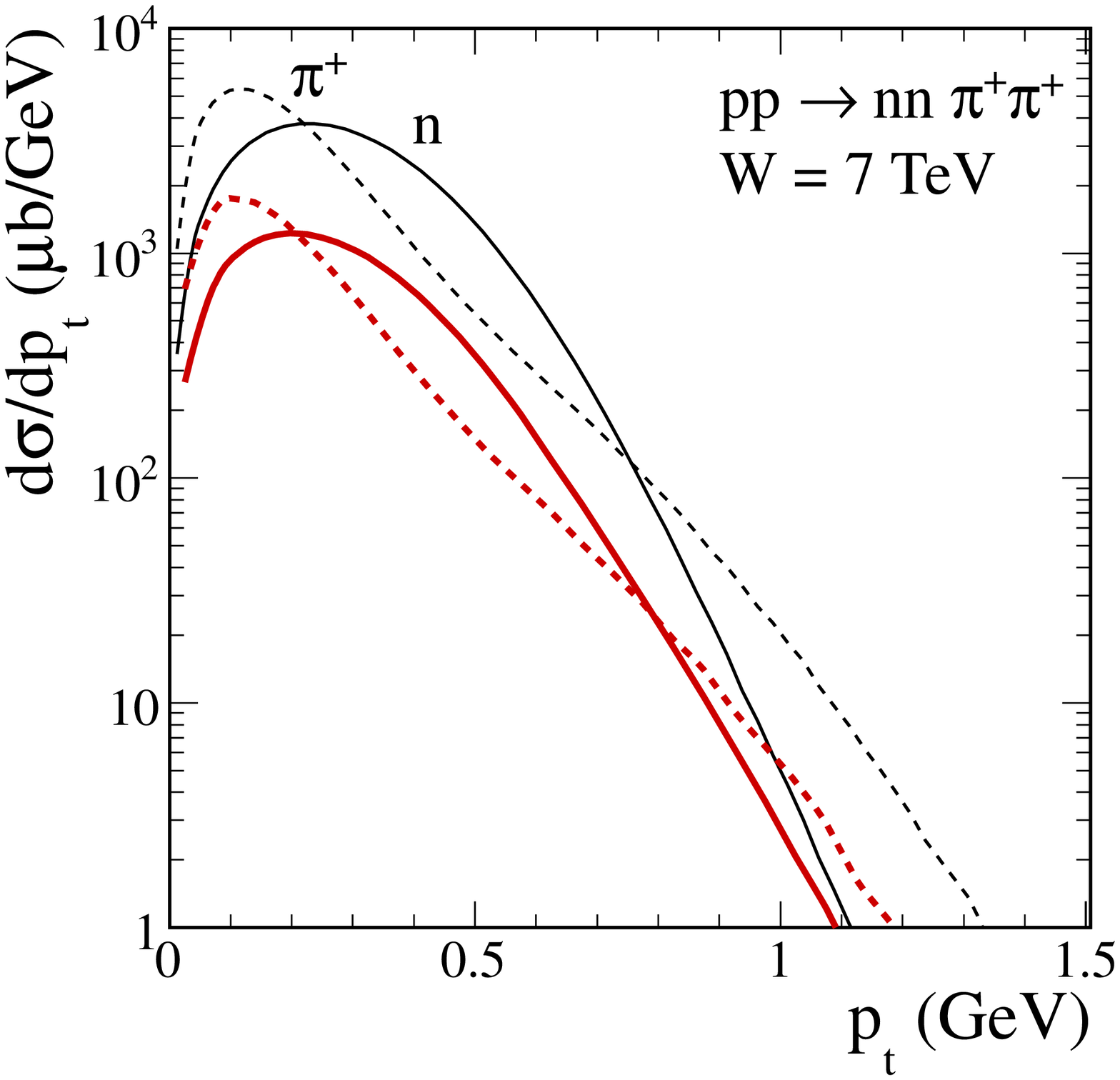}
  \caption{\label{fig:pt}
  \small
Differential cross section $d\sigma/dp_{t}$
for the $pp \to nn \pi^{+}\pi^{+}$ reaction
at $W$ = 7 TeV.
The solid and dotted lines correspond to the distribution 
in the transverse momentum of neutrons
and pions, respectively.
The lower curves correspond to calculations
with absorption effects.
}
\end{figure}

The distributions in the transverse momentum of neutrons and pions
are shown in Fig.\ref{fig:pt}.
The figure shows that the typical transverse momenta
are rather small but large enough to be measured. 
The distributions for neutrons are rather
similar to those for pions.

\begin{figure}[!h]
\includegraphics[width = 0.4\textwidth]{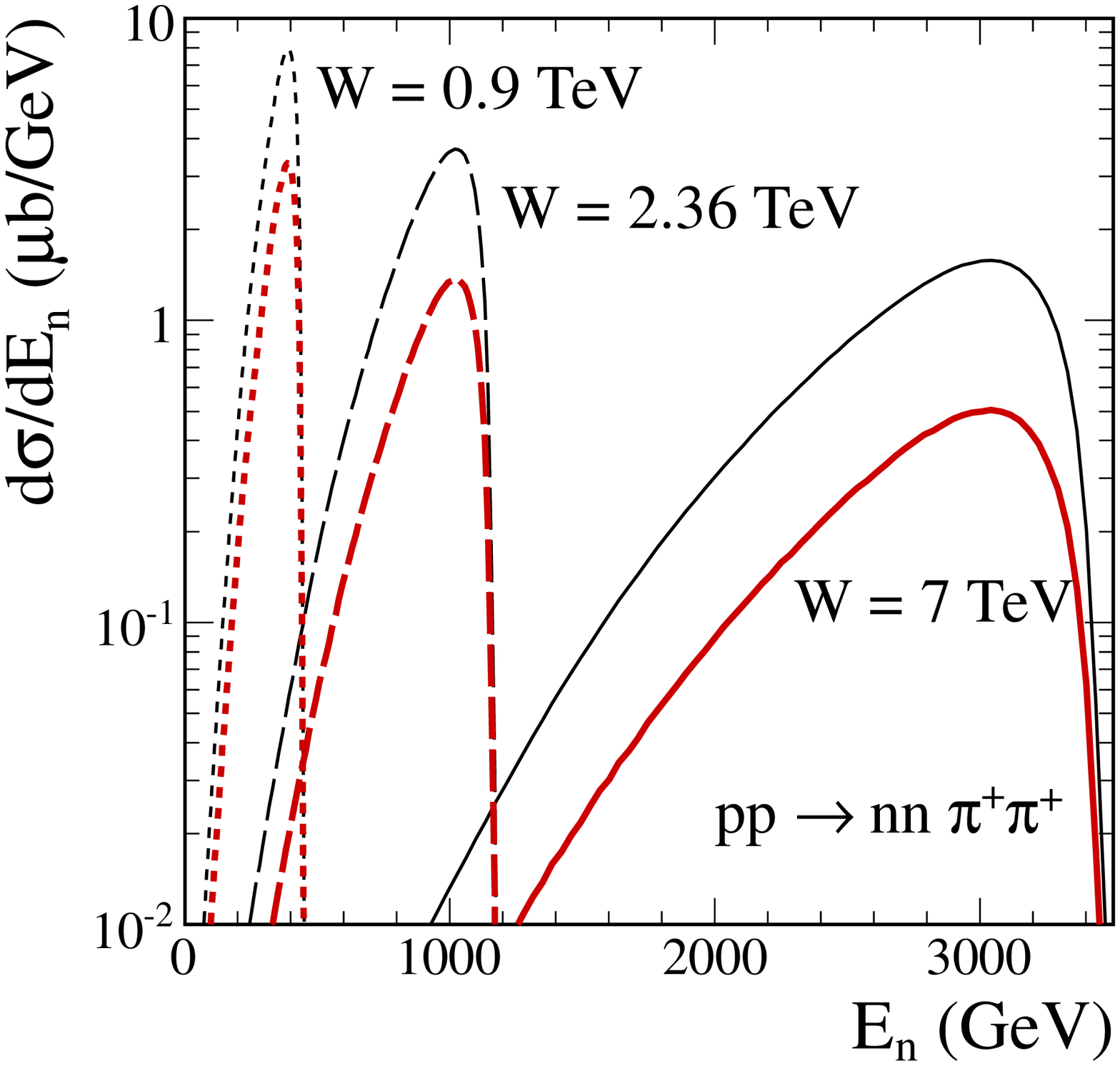}
\includegraphics[width = 0.4\textwidth]{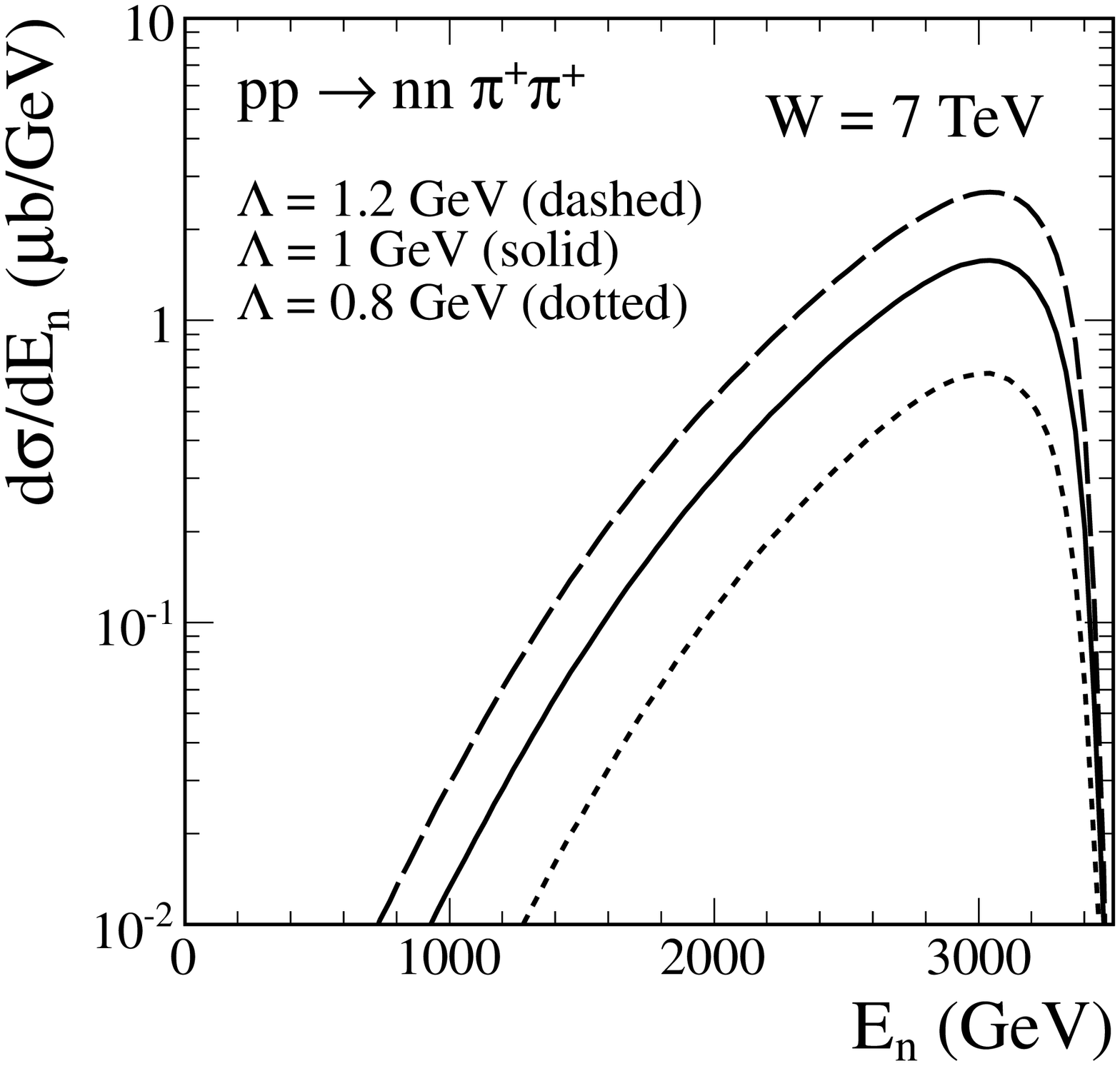}
  \caption{\label{fig:En}
  \small
Differential cross section $d\sigma/dE_{n}$
for the $pp \to nn \pi^{+}\pi^{+}$ reaction
at $W$ = 0.9, 2.36, 7 TeV (left panel).
The lower curves correspond to calculations
with absorption effects.
Right panel shows "bare" cross section obtained with different values
of the form factor parameter $\Lambda = 0.8$ GeV (dotted line),
$\Lambda = 1$ GeV (solid line) and $\Lambda = 1.2$ GeV (dashed line)
at $W$ = 7 TeV.
}
\end{figure}

The energy distributions of neutrons
are presented in Fig.\ref{fig:En}.
Generally the larger collision energy the larger energy
of outgoing neutrons. When combined with the previous plot
it becomes clear that the neutrons are produced
at very small polar angles (large pseudorapidities) and can be
measured by the ZDC's (see also Fig.\ref{fig:pseudorapidity_7000}).

\begin{figure}[!h]
\includegraphics[width = 0.4\textwidth]{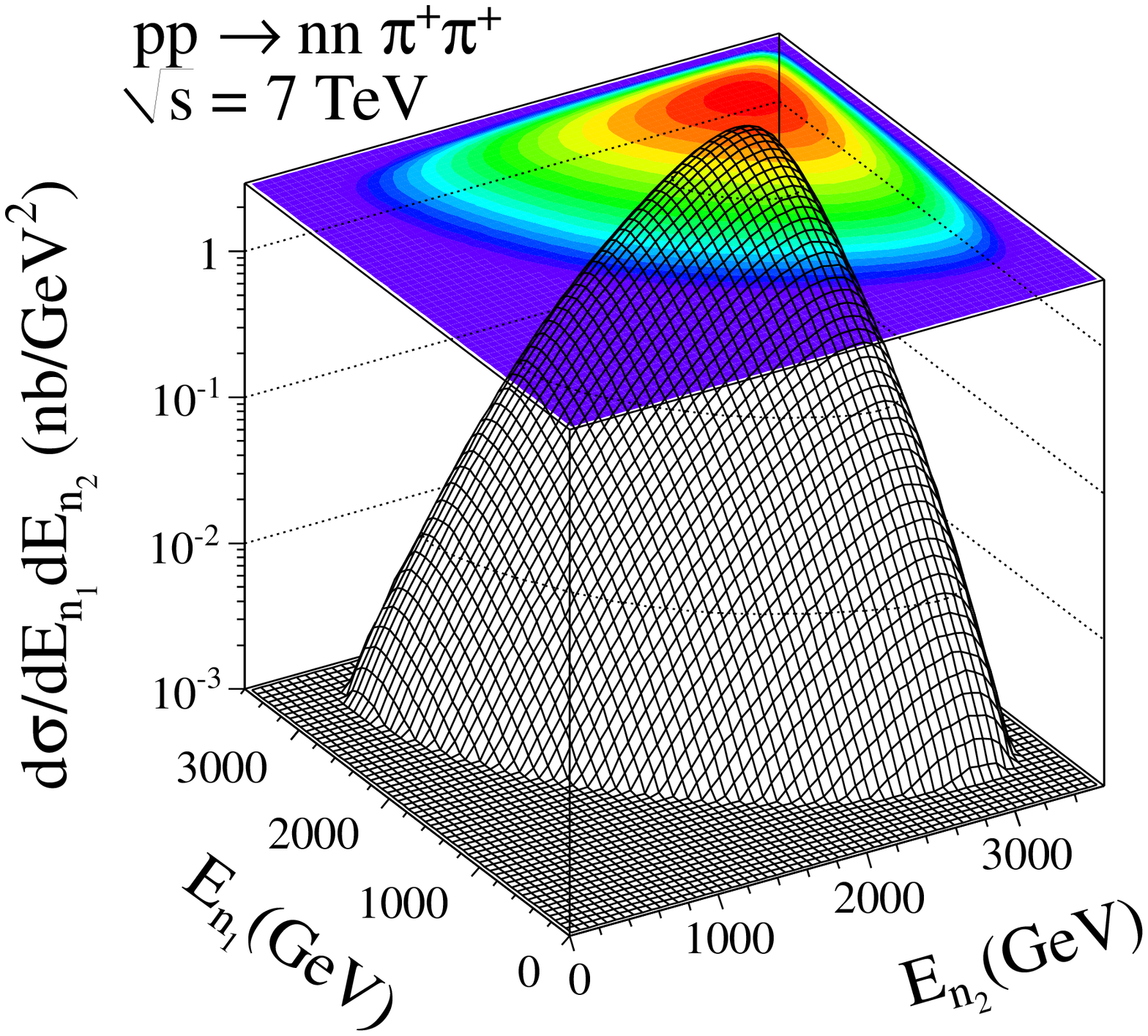}
  \caption{\label{fig:En_2d_7}
  \small
Differential cross section $d\sigma/dE_{n_{1}}dE_{n_{2}}$
for the $pp \to nn \pi^{+}\pi^{+}$ reaction at $W$ = 7 TeV.
}
\end{figure}

In Fig.\ref{fig:En_2d_7} we show two-dimensional correlations between
energies of both neutrons measured in both ZDC's.
The figure shows that the energies of both neutrons are
almost not correlated i.e. the shape (not the normalization)
of $d\sigma/dE_{n_{1}}$ ($d\sigma/dE_{n_{2}}$)
is almost independent of $E_{n_{2}}$ ($E_{n_{1}}$).
There should be no problem in measuring
energy spectra of neutrons on both sides as well as 
two-dimensional correlations in $(E_{n_{1}},E_{n_{2}})$.

\begin{figure}[!h]
\includegraphics[width = 0.4\textwidth]{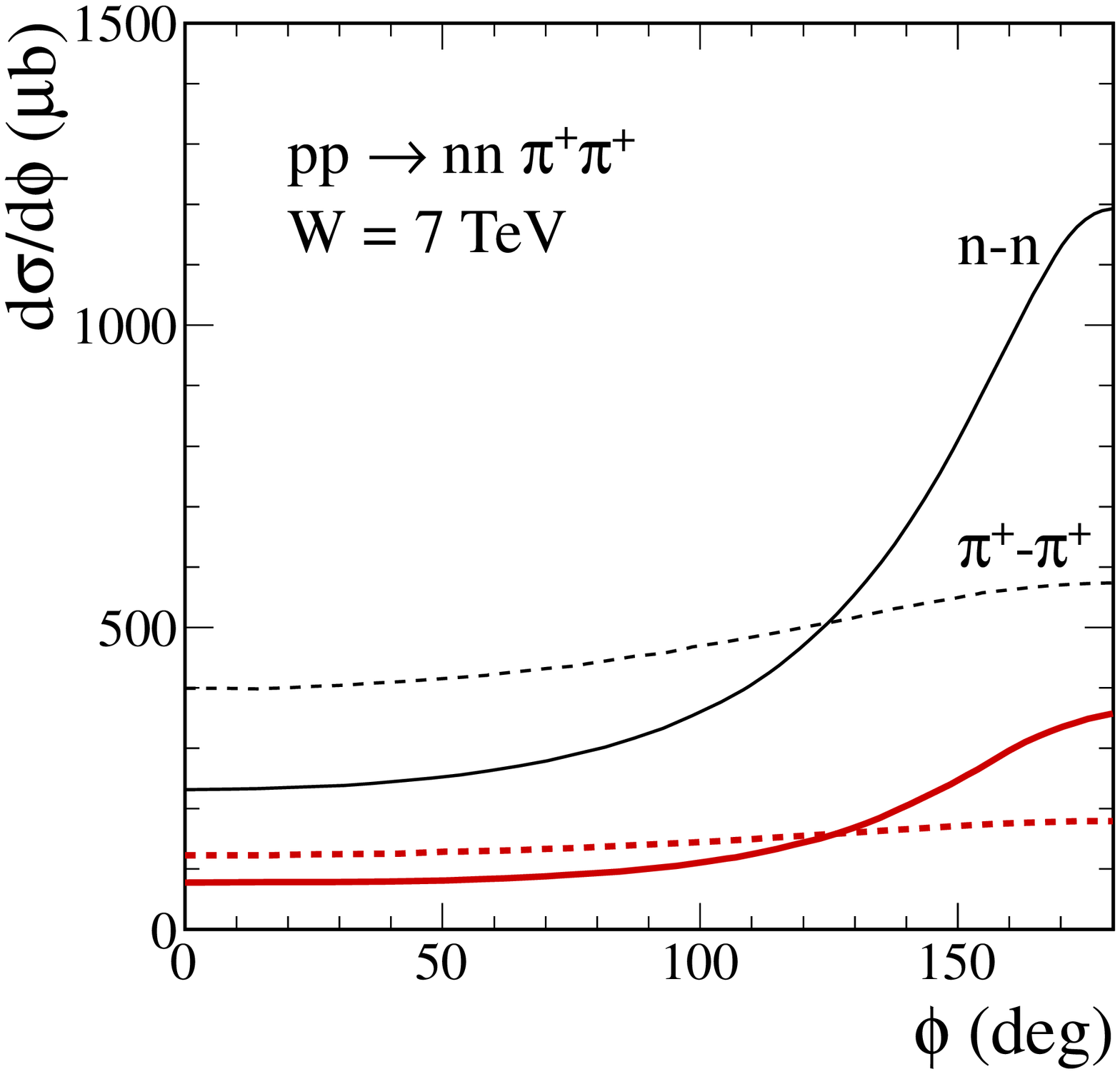}
  \caption{\label{fig:phi}
  \small
Azimuthal angle correlations between neutrons and beetwen pions
for the $pp \to nn \pi^{+}\pi^{+}$ reaction
at $W$ = 7 TeV.
The lower curves correspond to calculations
with absorption effects included.
}
\end{figure}

Finally in Fig.\ref{fig:phi} we present the distributions
in azimuthal angle $\phi$ between the transverse momenta
of the outgoing neutrons (pions).
Clearly a preference of back-to-back emissions can be seen.
The measurement of azimuthal correlations of neutrons 
will be not easy with first version of ZDC's as only 
horizontal position can be measured. Still
corellations of horizontal hit positions on both sides
could be interesting. A new correlation observable,
taking into account possibilities of the apparatus, 
should be proposed.
In contrast the two $\pi^{+}$'s are almost not correlated
in azimuthal angle.
However, such a distribution may be not easy to measure.

We have shown that at present the reaction under consideration
can be strictly measured only in a rather limited
part of the phase space (midrapidities of pions)
where the cross section is rather
small and where the double charged reggeon mechanism dominates. 
In Table \ref{tab:sig_tot_cross_cuts} 
we have collected the cross sections in nb for different
experiments at LHC and RHIC. At LHC where the separation of the
double-reggeon exchange mechanism is possible the cross section
is rather small of the order of a fraction of pb.
At RHIC the cross section with experimental cuts
should be easily measurable as it is of the order of a fraction of nb.

\begin{table}
\caption{Cross section (no absorption effects)
with different experimental 
cuts on $p_{t,\pi}$, $\eta_{\pi}$ and $\eta_{n}$.}
\label{tab:sig_tot_cross_cuts}
\begin{center}
\begin{tabular}{|c||c|c|c|c|c|c|}
\hline
      &$W$ (TeV) & $p_{t,\pi}>$ &  $|\eta_{\pi}|<$& $|\eta_{n}|_{_{ZDC}}>$ & $\sigma$ (nb)\\
\hline
ALICE & 7   & 0.15  & 0.9 &  8.7 & 6.3$\times$10$^{-5}$ \\
ALICE & 7   & 0.15  & 1.2 &  8.7 & 1.2$\times$10$^{-4}$ \\
ATLAS & 7   & 0.5   & 2.5 &  8.3 & 4.9$\times$10$^{-4}$ \\  
CMS   & 7   & 0.75  & 2.4 &  8.5 & 4.5$\times$10$^{-4}$ \\
RHIC  & 0.5 & 0.2   & 1   &  $-$ & 2.0$\times$10$^{-2}$ \\
\hline
\end{tabular}
\end{center}
\end{table}

\section{Conclusions}

We have estimated cross sections and calculated several differential observables
for the exclusive $p p \to n n \pi^+ \pi^+$ reaction.
Because our parameters are extracted from the analysis of known
two-body reactions we expect that our prodictions of the cross section
are fairly precise inspite of the complications of the reaction mechanism.
The full amplitude was parametrized in terms of
leading pomeron and subleading reggeon trajectories.
We have consider 3 classes of diagrams.
The first class gives the largest contribution
but concentrated at forward or backward pion directions.
There are also diagrams with double charged exchanges
with subleading reggeons $\rho^{+}$ and $a_{2}^{+}$.
Although the cross section
for these contributions is rather small, it is concentrated at midrapidities
of pions where the cross section can be easily measured.
The double-exchange reggeons processes
can be separated out in the two-dimensional space of rapidities of both pions
or in the distribution of the pion rapidity difference.

Large cross sections have been obtained,
even bigger than for the $p p \to p p \pi^+ \pi^-$ reaction \cite{LS10}.
Several mechanisms contribute to the cross section,
which leads to an enhancement of the cross section due to
interference effects.
These interference effects cause that the extraction
of the elastic $\pi^{+}\pi^{+}$ cross section
as proposed recently \cite{SRPM10} seems in practice rather impossible.

The specifity of the reaction is that both neutrons and pions
are emitted in very forward/backward directions,
producing a huge rapidity gap at midrapidities.
While the neutrons could be measured by the ZDC's,
the identification of pions may be difficult.
We think that the measurement of both neutrons and observation
of large rapidity gap is a very good signature of the considered reaction.
We expect the cross section for the $nn \pi^+ \pi^+ \pi^0$, $nn \pi^+ \pi^+ \pi^0 \pi^0$, etc.,
which could destroy rapidity gaps,
to be smaller but a relevant estimates need to be done.
In addition for events with larger number of pions 
the rapidity gap would be destroyed. Therefore the formally kinematically incomplete
measurement of two neutrons only could be relatively precise.
We have found that the neutrons measured in ZDC's seem to be
almost uncorrelated in energies.

We have made predictions for azimuthal angle correlations of outgoing
neutrons. Such distribution should be possible to measure in a future.
At present at CMS only horizontal position can be measured.
We have predicted back-to-back correlations with a sizeable diffusion. 

We have included elastic rescattering effects in a way used
recently for the three body processes. These effects lead
to a substantial damping of the cross section. The bigger energy the larger
the effect of damping. Other processes (e.g. inelastic intermediate states
or final state $\pi^{+} n$ interactions) 
could lead to additional damping. At present there is no full understanding
of the absorption effects.
A future experiment could provide new data to be analyzed and
could shed new light on absorption effects which are essential
for understanding exclusive processes, even such
important ones as exclusive production of the Higgs boson.

In the light of our analysis it becomes clear that 
extraction of the elastic $\pi^{+} \pi^{+}$ cross section
seems impossible, due to interference of several processes discussed in our paper.
We did not find any corner of the phase space
where the relevant diagram dominates.

There is an attempt to install forward
shower counters in the LHC tunnel.
Most probably they will not be able to measure
energy of the pions but they can signal some 
activity there. We expect that "some activity"
will mean, with a high probability, just one $\pi^+$
on one side and the other $\pi^+$ on the other side.

\section{Appendix}

\begin{figure}[!h]    %
 \includegraphics[width=0.18\textwidth]{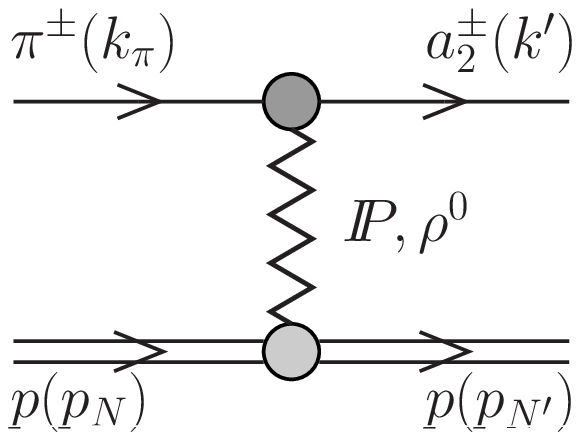}
 \includegraphics[width=0.18\textwidth]{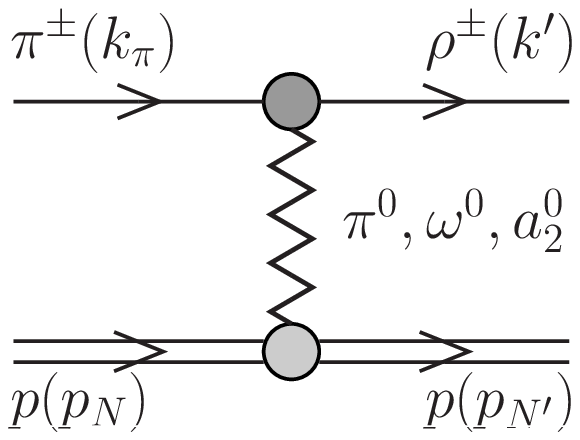}
 \includegraphics[width=0.18\textwidth]{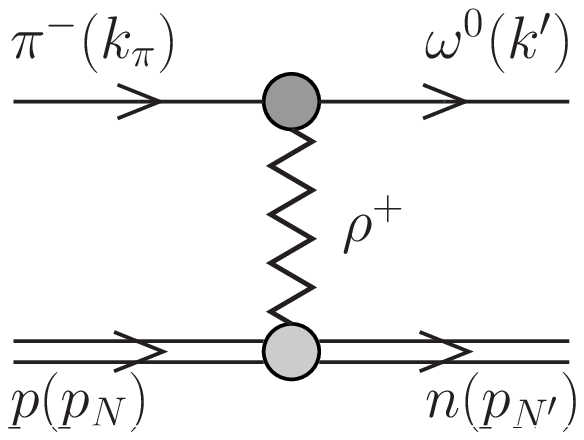}
 \includegraphics[width=0.18\textwidth]{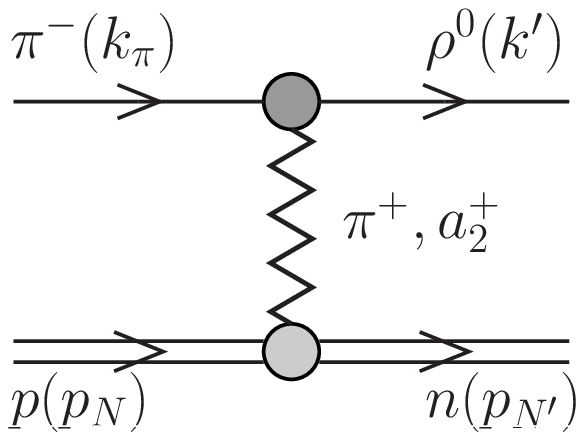}
  \caption{\label{fig:diagrams_abc}
  \small
Diagrams for various exchanges 
in $\pi p$ collisions.
}
\end{figure}
 
The $\rho$-meson/reggeon and $a_{2}$-meson/reggeon exchanges 
are known to have not only the nucleon spin-conserving part  
but also the dominant nucleon spin-flip component
while the $\omega$-meson/reggeon exchange to nucleons is mainly spin-conserving.
We write the amplitude for the reggeon exchanges (see Fig.\ref{fig:diagrams_abc})
in the following compact
phenomenological form:
\footnote{For the case of the $\pi^{-} p \to \omega^{0} n$, $\pi^{-} p \to \rho^{0} n$ 
and $\pi^{-} p \to a_{2}^{0} n$ reactions 
the amplitude should be multiplied by
$\sqrt{2}$ which is relatad to isospin Clebsch-Gordon coefficient.} 
\begin{eqnarray}
{\cal M}_{\lambda_{N} \to \lambda_{N'}, \lambda_{M}}^{reggeon-exch.}(s,t) &=&
\frac{\sqrt{-(t - t_{min})}}{M_0}
\left(\frac{-(t - t_{min})}{4 m_N^{2}}\right)^{|\lambda_{N'}-\lambda_{N}|/2} 
r_{T}^{i\;|\lambda_{n'}-\lambda_{n}|}\nonumber \\
&\times &\eta_{i}\; s \;C_{i}^r 
\left( \frac{s}{s_0}\right)^{\alpha_{i}(t)-1}
\exp \left( {\frac{B_{MN}}{2} (t-t_{min})}\right) \;\delta_{|\lambda_{M}|1} \;,
\label{amp_2to2_rho}
\end{eqnarray} 
and the pion exchange amplitude as
\begin{eqnarray}
{\cal M}_{\lambda_{N} \to \lambda_{N'}, \lambda_{\rho}}^{\pi-exch.}(s,t) &=&
g_{\pi N N}\;   F_{\pi NN}(t)\; 
\bar{u}(p_{N'},\lambda_{N'}) \mathrm{i} \gamma_{5} u(p_{N},\lambda_{N})\nonumber \\
&\times & (k_{\pi}^{\mu}+q^{\mu}) 
\epsilon^{*}_{\mu}(k', \lambda_{\rho})\;
\dfrac{\mathrm{i}}{t -m_{\pi}^{2}}\;
g_{\rho \pi \pi}\;  F_{\rho \pi \pi}(t) \;
\left(\frac{s}{s_0}\right)^{\alpha_{\pi}(t)} \;.
\label{amp_2to2_pi}
\end{eqnarray} 

Above the $\sqrt{-(t - t_{min})}/M_0$ factor 
is due to the meson spin-flip 
(in the $\pi \to \omega$, $\pi \to \rho$ and $\pi \to a_2$ transitions), 
$M_0$ is a reference scale factor taken here
$M_0$ = 1 GeV (which is used here to have the same units
for the coupling constants). 
The double spin-flip components do not interfere
with the spin-conserving ones and can be calculated separately.
Here we have introduced one more phenomenological (dimensionless)
parameter $r_{T}^{i}$ which describes coupling for the spin-flip components.
It is known to be of 
$r_{T}^{\rho}  = 7.5$, 
$r_{T}^{a_{2}} \simeq 6.14$,
$r_{T}^{\omega}\simeq 0.17$ \cite{KS76} and
$r_{T}^{\rho}  \simeq 8$, 
$r_{T}^{a_{2}} \simeq 4.7$,
$r_{T}^{\omega}\simeq 0.9$ \cite{IW77}.
In the present calculations we take $r_{T}^{\rho}$ = 7.5, 
$r_{T}^{a_{2}}$ = 6 and $r_{T}^{\omega}$ = 0.
The coupling constant $g_{\rho \pi \pi}$ is taken as
$g_{\rho \pi \pi}^2 / 4 \pi$ = 2.6.
The form factors are parametrized as $F(t) = \exp\left((t-m_{\pi}^{2})/\Lambda^{2}\right)$.
We improve the parameterization of the amplitude 
(\ref{amp_2to2_pi})
by multiplying by the factor $(s/s_0)^{\alpha_{\pi}(t)}$,
where $\alpha_{\pi}(t)=\alpha_{\pi}'(t-m_{\pi}^{2})$ 
is the pion Regge trajectory with the slope of trajectory
$\alpha_{\pi}'=1$ GeV$^{-2}$.

We adjust the $C_{i}^r$ (where $i$ = $I\!\!P, \rho, \omega, a_{2}$) coupling
constants to the world experimental data 
often obtained from partial wave analysis
in the three-pion system.
The effective normalization constants for the auxiliary reactions
are related to those in the $NN$ scattering and 
the $g_{\pi \to a_{2}, \rho, \omega}^{i}$ coupling constants we need in our problem as:
\begin{eqnarray}
C_{i}^r = \sqrt{C_{i}^{NN}} \cdot g_{\pi \to j}^{i}\;.
\end{eqnarray}
Since $C_{i}^{NN}$ are known from phenomenology 
(Table \ref{tab:parameters}),
$g_{\pi \to j}^{i}$ can be obtained from our fits:
$g_{\pi \to a_{2}}^{I\!\!P}$ = 1.4 GeV$^{-1}$,
$g_{\pi \to a_{2}}^{\rho}$ = $g_{\pi \to \rho}^{a_{2}}$ = 22 GeV$^{-1}$ and 
$g_{\pi \to \rho}^{\omega}$ = $g_{\pi \to \omega}^{\rho}$ = 4 GeV$^{-1}$.
%

\begin{figure}[!h]
a) \includegraphics[width = 0.4\textwidth]{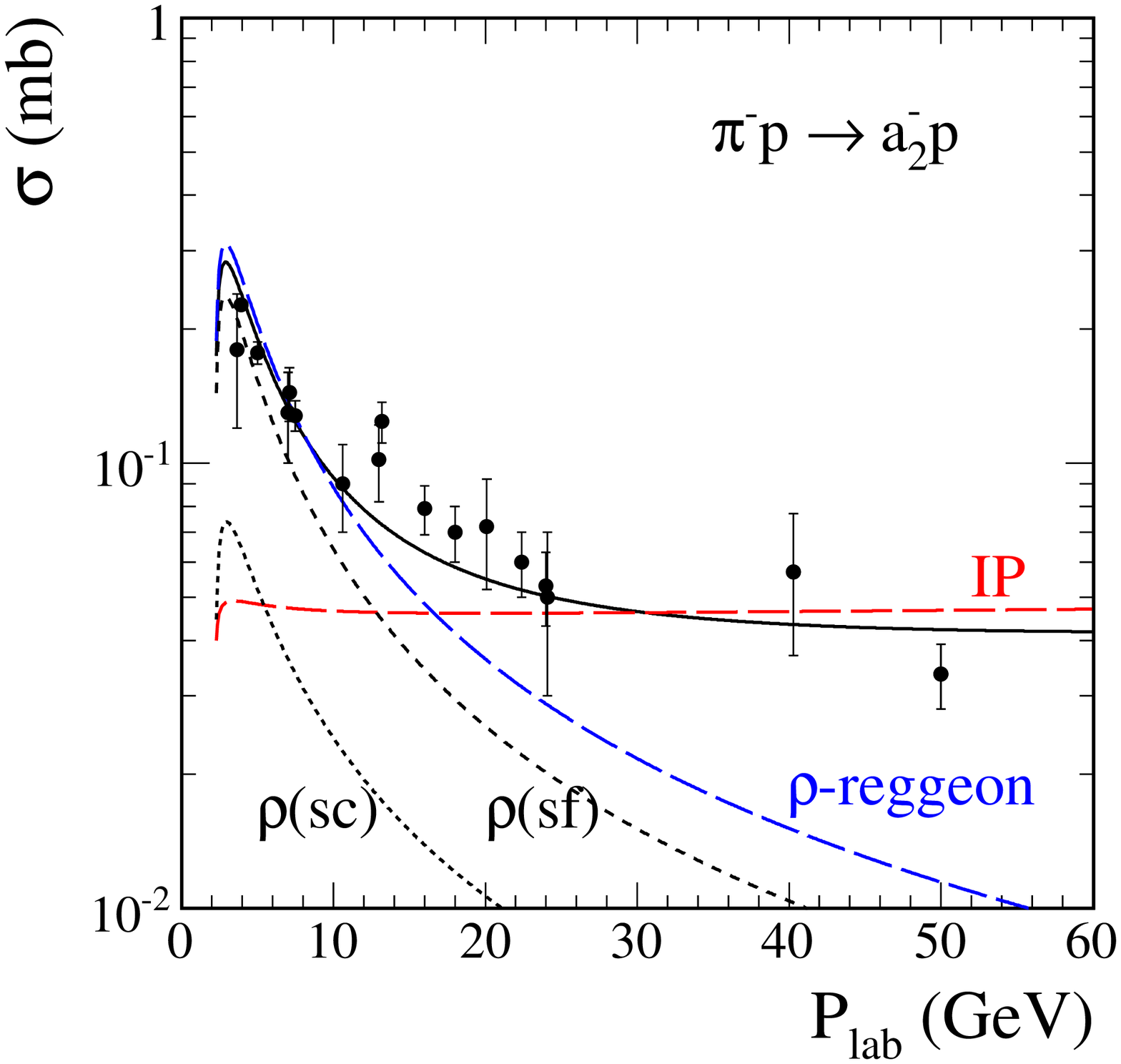}
b) \includegraphics[width = 0.4\textwidth]{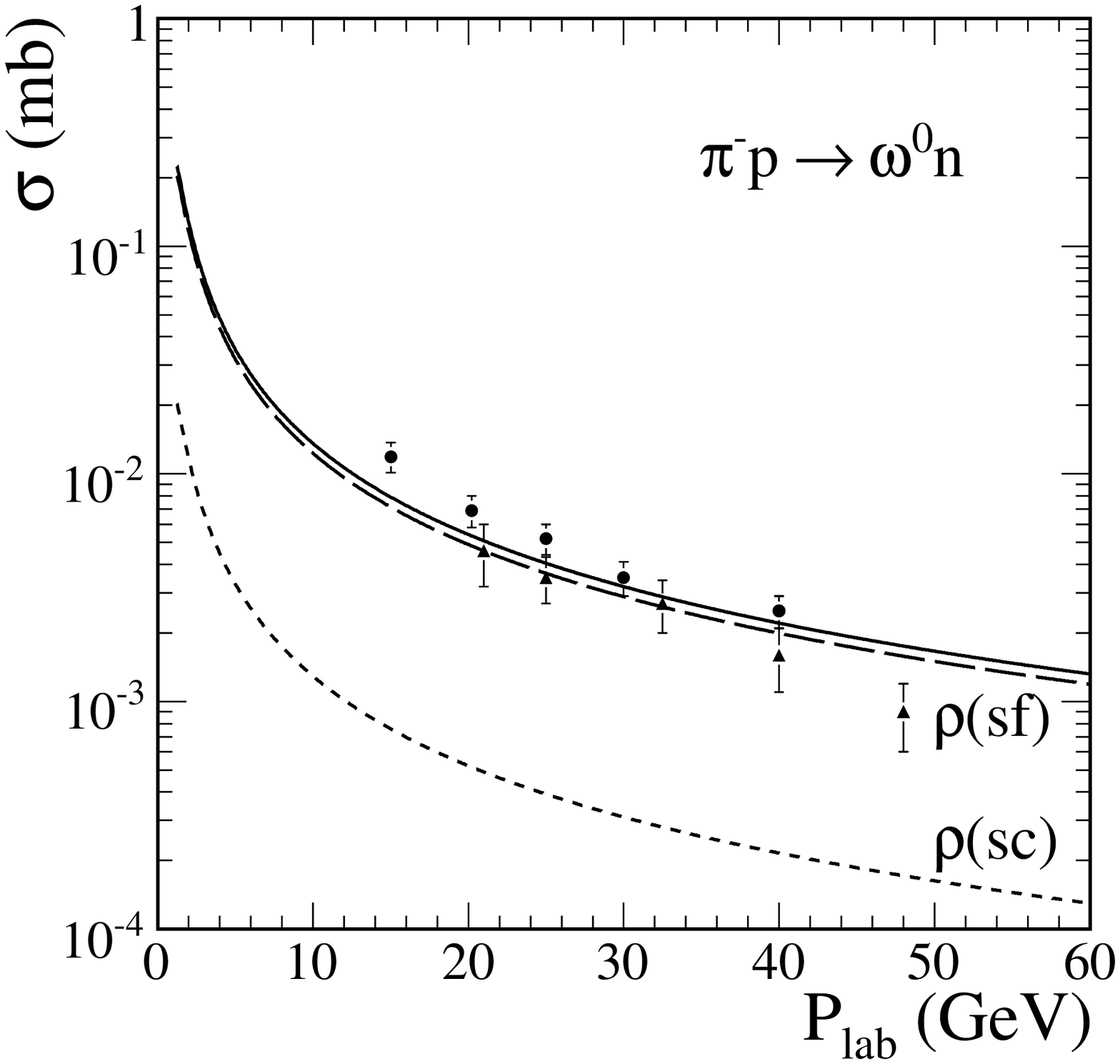}\\
c) \includegraphics[width = 0.4\textwidth]{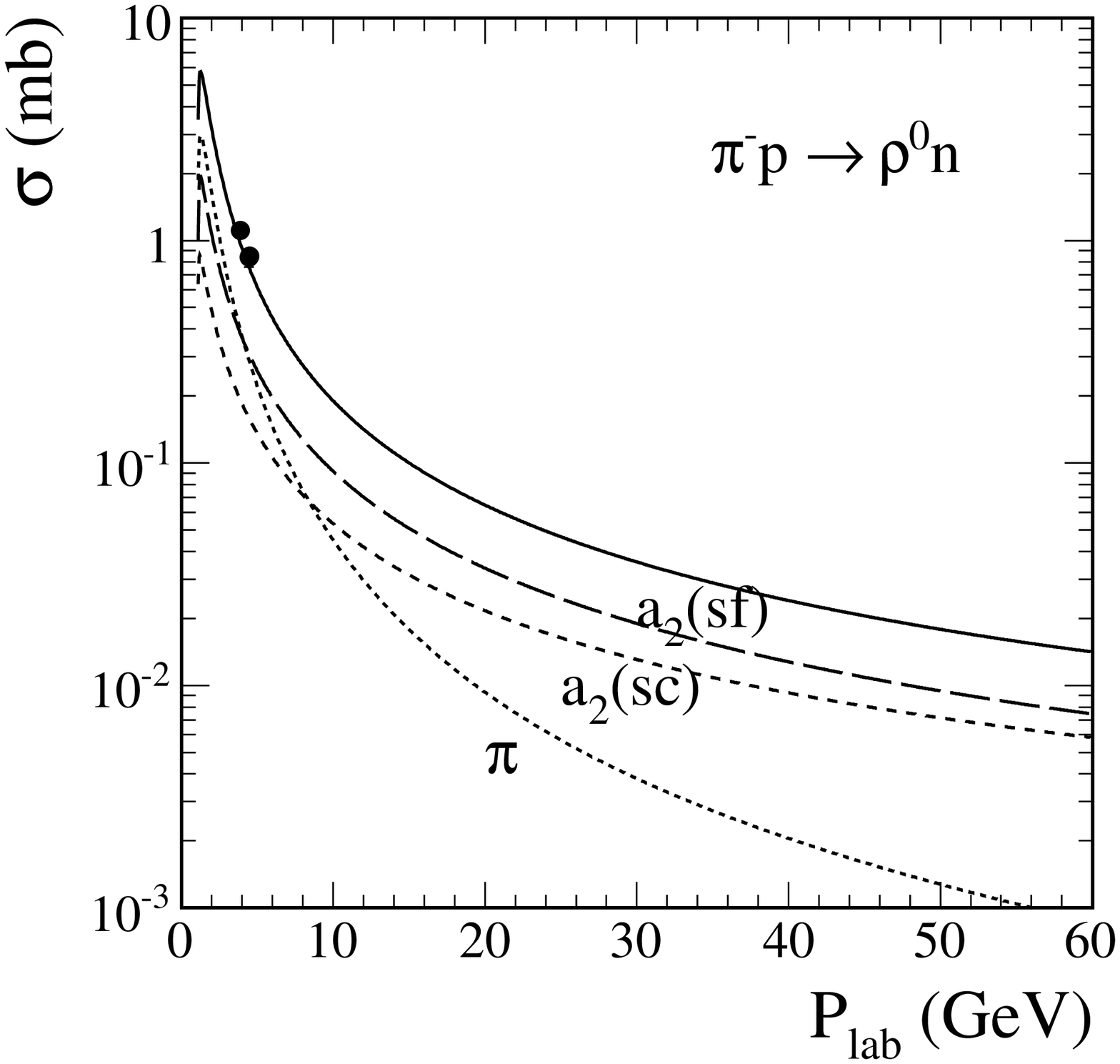}
d) \includegraphics[width = 0.4\textwidth]{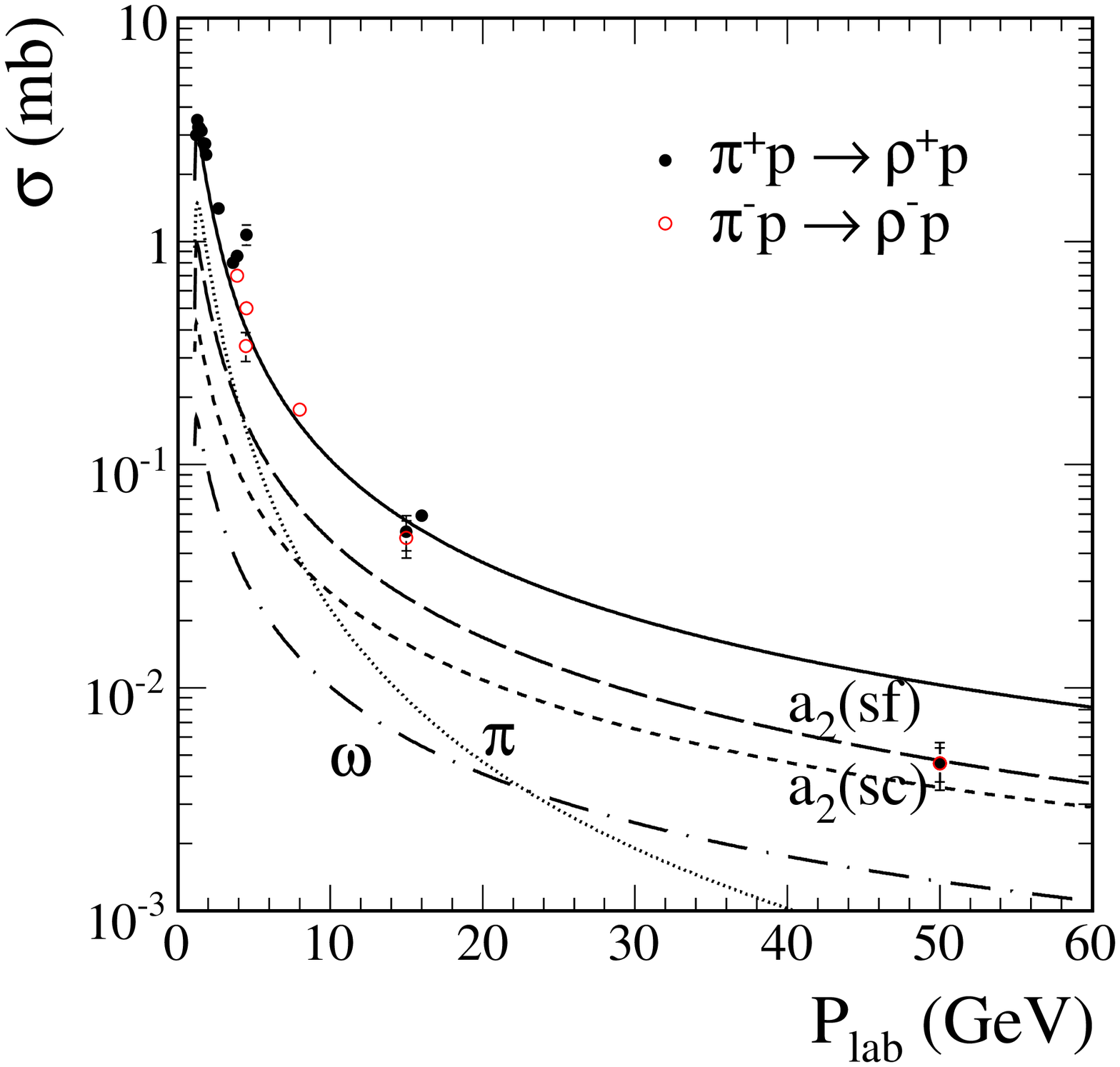}
  \caption{\label{fig:sig_pim_a2m}
  \small
The integrated cross section
for the $\pi^- p \to a_2^- p$
(the experimental data are taken from \cite{a2mp,Delfosse81}),
$\pi^- p \to \omega^0 n$ \cite{omen},
$\pi^- p \to \rho^0 n$ \cite{Kartamyshev,HaberAlekseeva},
$\pi^+ p \to \rho^+ p$ \cite{Delfosse81,HaberAlekseeva,rhopp,Pratt} and
$\pi^- p \to \rho^- p$ \cite{Delfosse81,Kartamyshev,HaberAlekseeva,Pratt,Bugg82} reactions
as a function of the incident-beam momenta $P_{lab}$.
}
\end{figure}

In Fig.\ref{fig:sig_pim_a2m} we show the total cross section 
for the $\pi^- p \to a_2^- p$,
$\pi^- p \to \omega^0 n$,
$\pi^{-} p \to \rho^{0} n$
and $\pi^{\pm} p \to \rho^{\pm} p$ reactions as
a function of the incident-beam momenta $P_{lab}$. 
Our fit is shown by the solid line.
In the panel a) ($\pi^{-}p \to a_2^{-} p$ reaction) we show individual contributions
of $\rho$ and pomeron exchanges. The pomeron exchange dominates at high energies 
whereas the $\rho$ exchange at small energies. This separation of mechanisms
allows to extract two independent coupling constants. 
We show also spin-conserving and spin-flip amplitudes separately.
In panel b) we show our fit for the $\pi^{-} p \to \omega^{0} n$.
Here only $\rho$ exchange is possible.
In panel c) ($\pi^{-} p \to \rho^{0} n$ reaction)
we show contributions for charged pion exchange (parameters fixed from phenomenology)
and  $a_2$ exchange (parameters found from the analysis of
the $\pi^{-}p \to a_2^{-} p$ (see panel a))).
Finally in panel d) ($\pi^{\pm} p \to \rho^{\pm} p$ reactions) we show
contributions for neutral pion exchange, $a_2$ exchange and
$\omega$ exchange (relevant coupling constant found from the analysis of the $\pi^{-} p \to \omega^{0} n$
reaction (see panel b))).

Having fixed the parameters
we can proceed to our four-body $p p \to n n \pi^+ \pi^+$ reaction.

\vspace{0.5cm}
{\bf Acknowledgments}

We are indebted to Michael Murray for an interesting discussion
on a possibility of a measurement of the discussed reaction and
Wolfgang Sch\"afer for a discussion of the reaction mechanisms.
This study was partially supported by the Polish grant
of MNiSW No. N202 249235.


\end{document}